\documentclass[]{aa}
\usepackage{graphicx, natbib, longtable, lscape, amssymb}
\usepackage[english]{babel}
\usepackage{lscape}
\usepackage{rotating}

\begin{document}

\title{Monte Carlo  simulations of post-common-envelope white  dwarf +
  main  sequence  binaries:  comparison  with the  SDSS  DR7  observed
  sample}

\author{Judit Camacho\inst{1,2}, 
        Santiago Torres\inst{1,2}, 
        Enrique Garc\'{\i}a--Berro\inst{1,2}, 
        M\'{o}nica Zorotovic\inst{3}, 
        Matthias R. Schreiber\inst{3}, 
        Alberto Rebassa--Mansergas\inst{4},
        Ada Nebot G\'omez--Mor\'an\inst{5} \and
        Boris T. G\"ansicke\inst{6}}

\institute{Departament de F\'\i sica Aplicada, 
           Universitat Polit\`ecnica de Catalunya,
           c/Esteve Terrades 5,         
           08860 Castelldefels, Spain 
           \and
           Institute for Space Studies of Catalonia,
           c/Gran Capit\`a 2--4, Edif. Nexus 104, 
           08034 Barcelona, Spain 
           \and 
           Departamento de F\'{\i}sica y Astronom\'{\i}a,
           Universidad de Vapara\'{\i}so,
           Avda. Gran Breta\~na 1111,
           Valpara\'{\i}so, Chile 
           \and
           Kavli Institute for Astronomy and Astrophysics, 
	   Peking University, 
	   Beijing 100871, 
	   China 
           \and
           Observatoire Astronomique de Strasbourg, 
           Universit\'e de Strasbourg, 
           CNRS, UMR 7550, 11 rue de l'Universit\'e, 67000, 
           Strasbourg, 
           France 
           \and
           Department of Physics, 
           University of Warwick, 
           Coventry CV4 7AL, UK}

\date{\today}

\titlerunning{Simulations of the WD+MS PCEBs in the SDSS DR7}
\authorrunning{Camacho et al. }

\offprints{E. Garc{\'\i}a--Berro}


\abstract
{Detached white  dwarf + main  sequence (WD+MS) systems  represent the
  simplest population of post-common  envelope binaries (PCEBs). Since
  the  ensemble  properties  of   this  population  carries  important
  information about  the characteristics  of the  common-envelope (CE)
  phase,  it  deserves  close   scrutiny.   However,  most  population
  synthesis studies do not fully take  into account the effects of the
  observational selection biases  of the samples used  to compare with
  the theoretical simulations.}
{Here  we  present the  results  of  a  set  of detailed  Monte  Carlo
  simulations of the population of WD+MS binaries in the Sloan Digital
  Sky Survey (SDSS) Data Release 7.}
{We used up-to-date stellar  evolutionary models, a complete treatment
  of the Roche lobe overflow episode, and a full implementation of the
  orbital evolution of the binary  systems. Moreover, in our treatment
  we  took into  account  the  selection criteria  and  all the  known
  observational biases.}
{Our  population  synthesis study  allowed  us  to make  a  meaningful
  comparison with the available  observational data. In particular, we
  examined the  CE efficiency,  the possible contribution  of internal
  energy, and the initial mass ratio distribution (IMRD) of the binary
  systems.   We found  that  our simulations  correctly reproduce  the
  properties  of  the  observed   distribution  of  WD+MS  PCEBs.   In
  particular,  we  found  that   once  the  observational  biases  are
  carefully taken  into account,  the distribution of  orbital periods
  and of masses of the WD and MS stars can be correctly reproduced for
  several choices of the free parameters and different IMRDs, although
  models in  which a  moderate fraction ($\le  10\%$) of  the internal
  energy  is used  to eject  the CE  and in  which a  low value  of CE
  efficiency is used ($\le 0.3$)  seem to fit better the observational
  data.    We  also   found  that   systems  with   He-core  WDs   are
  over-represented in the observed sample, due to selection effects.}
{Although  our  study represents  an  important  step forward  in  the
  modeling  of  the  population  of  WD+MS  PCEBs,  the  still  scarce
  observational data  precludes to derive without  ambiguity a precise
  value of the several free parameters used to compute the CE phase or
  to ascertain which is the correct IMRD.}


\maketitle


\section{Introduction}
\label{intro}

Close-compact  binaries  are  at  the  heart  of  several  interesting
phenomena  in  our  Galaxy  and in  other  galaxies.   In  particular,
cataclysmic variables,  low mass  X-ray binaries or  double degenerate
white dwarf  (WD) binaries -- just  to mention the most  important and
well-studied ones  -- are systems  that not only deserve  attention by
themselves,  but  also  because their  statistical  distributions  are
crucial to understand the underlying physics of the evolution during a
common envelope episode.  Actually, the vast majority of close-compact
binaries are formed through at least one CE episode. This phase occurs
when the  more massive  star, hereafter the  primary, fills  its Roche
lobe  during the  first  giant  branch (FGB)  or  when  it climbs  the
asymptotic  giant   branch  (AGB).   The  mass   transfer  episode  is
dynamically unstable  and the envelope  of the giant star  engulfs the
less massive star, i.e. the  secondary, forming a common envelope (CE)
around both the core of the  primary (the future compact star) and the
secondary  star.   Drag forces  transfer  orbital  energy and  angular
momentum  to the  envelope,  leading  to a  dramatic  decrease of  the
orbital separation,  and to  the ejection  of the  CE.  If  the system
survives the CE  phase, the outcome is a post-CE  binary (PCEB) formed
by  a compact  object and  the main  sequence (MS)  companion with  an
orbital period separation much smaller  than that of the original main
sequence binary system. The PCEBs studied  in detail in this paper are
those in which the compact object is a WD.

Even though the basic concepts of  the evolution during a CE phase are
rather   simple,  the   details  are   still  far   from  being   well
understood. This is so because several complex physical processes play
an important role in the evolution  during the CE phase. For instance,
the spiral-in of the core of the primary and of the secondary, and the
ejection of the  envelope are not only a consequence  of the evolution
of the  core and  remaining layers  of the donor  star in  response to
rapid mass loss, but also tidal  forces and viscous dissipation in the
CE play  key roles. Moreover,  these physical processes occur  on very
different timescales  and on a wide  range of physical  scales  -- see
\citet{Taam_2010}    for   a    recent   review.     Consequently,   a
self-consistent   modeling  of   the   CE   phase  requires   detailed
hydrodynamical models  which are  not available  at the  present time,
although recent  progresses are  encouraging --  see \citet{Taam_2012}
and references  therein.  Hence, the  CE phase has  been traditionally
described using parametrized models.

There are two canonical formalisms to  treat the evolution during a CE
episode.  The most commonly used one, known as the $\alpha$ formalism,
assumes energy  conservation \citep{Webbink_1984,kool_1990, dewi2000}.
The second formalism is based  on angular momentum conservation and it
is known  as the $\gamma$ formalism  \citep{Nelemans_2005}. Within the
$\alpha$  formalism,  the  energy   transferred  to  the  envelope  is
parametrized  using   an  efficiency  parameter,   $\alpha_{\rm  CE}$.
Furthermore, the binding  energy of the envelope is  also modeled with
another free parameter, $\lambda$, which mainly depends on the mass of
the donor and on its evolutionary stage.  The most recent formulations
also include a  third parameter, $\alpha_{\rm int}$, which  is used to
measure  the  fraction of  the  internal  energy contributing  to  the
ejection of the  envelope.  We postpone a precise  definition of these
parameters  to Sect.~\ref{s-CEP},  but  we emphasize  here that  these
parameters are still poorly determined.  Thus, studying the population
of binaries that have undergone a CE episode is important because some
of their  characteristics, like  the distribution of  orbital periods,
primary and secondary masses, can be used to constrain their values.

Binary systems formed by a WD and a MS companion are intrinsically one
of the  most common, and  structurally simplest, population  of PCEBs.
Thus, the  statistical properties of  this population are  expected to
provide crucial observational inputs that are necessary to improve the
theory   of   CE   evolution   \citep{davisetal10-1,   Zorotovic_2010,
demarcoetal11-1, arm_2012}.   However, until now,  detailed population
synthesis  studies  have  failed  to effectively  constrain  the  free
parameters involved  in the  formulation of  the CE  phase, due  to an
utter  lack of  observational  data --  see e.g.   \citet{dekool92-1},
\citet{willems+kolb04-1},       \citet{politano+weiler07-1},       and
\citet{davisetal10-1}.   In particular,  it  has been  shown that  the
early sample of well-studied PCEBs is  not only small but, being drawn
mainly from ``blue'' quasar surveys, it is also heavily biased towards
young      systems       with      low-mass       secondary      stars
\citep{schreiber+gaensicke03-1}.        However,        the       SDSS
\citep{yorketal00-1,  abazajianetal09-1}  has  allowed to  identify  a
large    number     of    WD+MS     binaries    \citep{helleretal09-1,
rebassa-mansergasetal13-2},  and a  dedicated  radial velocity  survey
among them has provided the so far largest and most homogeneous sample
of   close   compact   binaries   with   available   orbital   periods
\citep{rebassa-mansergasetal08-1, nebotetal11-1}

Recently,  \citet{toonen+nelemans13-1}   presented  binary  population
models of  WD+MS PCEBs taking  into account some of  the observational
selection  effects  important for  the  SDSS  sample.  However,  while
representing a  nice step forward,  the conclusions that can  be drawn
from their  study is limited by  their assumption of a  constant value
for  the binding  energy  parameter  and by  not  taking into  account
possible  contributions  from  the   internal  energy  stored  in  the
envelope.  In  this  paper  we  describe the  results  of  a  detailed
population synthesis study of WD+MS  PCEBs in the Galaxy, modeling all
the observational selection effects  affecting the observed population
in the  well-characterized sample of  PCEBs detected in the  SDSS Data
Release (DR) 7.  A direct comparison of the simulated and the observed
sample  of PCEBs  is  performed  as well,  with  the  ultimate aim  of
constraining  the current  theories  of CE  evolution.   The paper  is
organized as  follows.  Sect.~\ref{MC} describes the  main ingredients
of our Monte Carlo simulator,  whereas in Sect.~\ref{bias} the filters
applied in  order to  take into account  the observational  biases are
discussed in depth.  The observed sample to which the simulations will
be    compared   is    presented   in    Sect.~\ref{s-sample},   while
Sect.~\ref{s-results} presents  the main  results of  our simulations,
followed by an  exhaustive analysis of the role played  by some of the
parameters  involved in  the CE  phase.  Finally,  Sect.~\ref{s-concl}
closes  the  paper  with  a  summary of  our  main  findings  and  our
concluding remarks.


\section{The simulated population of WD+MS PCEBs}
\label{MC}

We expanded an existing Monte Carlo code \citep{MC1, MC2} specifically
designed to study  the Galactic population of single WDs  to deal with
the population of binaries in which one  of the components is a WD. In
this section we  describe in detail the most  important ingredients of
our Monte Carlo simulator.

\subsection{The Monte Carlo simulator}
\label{general}

The basic ingredient of any Monte  Carlo code is a generator of random
variables distributed  according to  a given probability  density. The
simulations described  in this paper  were done using a  random number
generator  algorithm  \citep{James_1990}   which  provides  a  uniform
probability  density  within  the   interval  $(0,1)$  and  ensures  a
repetition period  of $\ga 10^{18}$,  which is virtually  infinite for
practical simulations. When Gaussian probability functions were needed
we used the Box-Muller algorithm as described in \cite{NRs}.

We randomly chose two numbers for the galactocentric polar coordinates
$(r,\theta)$ of each  synthetic star of the  entire stellar population
within approximately  5~kpc from  the Sun and  following the  SDSS DR7
spectroscopic plate directions  \citep{abazajianetal09-1}. The adopted
density  distribution followed  an exponential  law with  radial scale
length of 3.5~kpc. The $z$ coordinate was randomly chosen following an
exponential law with scale height $H=250$~pc. We assumed a fraction of
binaries of 50\% and we normalized  our simulated systems to the local
disk  mass  density  \citep{Holmberg_2000}.   Next we  drew  two  more
pseudo-random numbers: the first one for the mass on the MS, $M_1$, of
each simulated primary star --  according to the initial mass function
of \cite{Kroupa_1993}  -- and the  second for  the time at  which each
star was born -- assuming a  constant star formation rate. The adopted
age of  the Galactic disk  was 10~Gyr.   Since the initial  mass ratio
distribution  (IMRD) is  still a  controversial issue,  we used  three
different prescriptions  for it.   The first one  consisted in  a flat
distribution  $n(q)=1$, with  $q=M_{2}/M_{1}$  the  mass ratio,  where
$M_1$ and  $M_2$ are the  masses of  the primary and  secondary stars,
respectively. We  also considered  a distribution of  secondary masses
that depends inversely on the  mass ratio, $n(q)\propto q^{-1}$, and a
distribution proportional to  the mass ratio, $n(q)\propto  q$. In all
cases, we only took into account stars with masses ranging from $0.1\,
M_{\sun}$ to  $30\,M_{\sun}$.  In  addition, orbital  separations were
randomly  drawn according  to a  logarithmic probability  distribution
\citep{Nelemans_2001b},  $f(a)\propto \ln  a$ for  $3 \leq  a/R_{\sun}
\leq  10^6$.    Finally,  the  eccentricities  were   randomly  chosen
according to a thermal distribution \citep{Heggie_1975}, $g(e)=2e$ for
$0.0 \leq e \leq 0.9$.

Once the  masses of the  stars were known,  and the properties  of the
binary  system were  assigned  according to  the previously  explained
procedures, each of the components was  evolved. We did that using the
analytical   fits  to   detailed   stellar   evolutionary  tracks   of
\cite{hurleyetal00-1}, which provide full coverage of the entire range
of masses  of interest  from the zero-age  main sequence  (ZAMS) until
advanced stages of evolution. These  evolutionary fits provide all the
relevant  information   --  such   as  radii,   masses,  luminosities,
evolutionary timescales,\ldots -- but  the photometric properties (see
Sect.~\ref{s-photom}).   We note  that the  evolutionary sequences  of
\cite{hurleyetal00-1}  are   conservative.   Accordingly,   to  obtain
realistic simulations mass loss must  be included. We assumed that the
evolution  during  the  MS  phase was  conservative,  and  only  after
hydrogen  starts burning  in a  shell we  considered mass  losses. The
adopted mass-loss rate was that  of \citet{Reimers_1978}, for which we
assumed an efficiency $\eta=0.5$. On the AGB phase the prescription of
\citet{Vassiliadis_1993} was  used.  In  the case of  moderately close
binary systems  we also considered  a tidally enhanced  mass-loss rate
\citep{te1988}:

\begin{equation}
\dot M =\dot M_{\rm R}
\left[1+B_{\rm W} \max\left( \frac{1}{2},
\frac{R}{R_{\rm L}}\right)^{6}\right] 
\label{tidal}
\end{equation} 

\noindent where $M$ and $R$ are  the mass and star radius, $R_{\rm L}$
is the  Roche-lobe radius, $\dot  M_{\rm R}$ is the  standard Reimers'
mass-loss   rate,  and   $B_{\rm   W}$  is   the  enhanced   mass-loss
parameter. As  it will be shown  below, we analyzed several  models in
which $B_{\rm W}$ varies from 0 to $10^4$. Angular momentum losses due
to  magnetic  braking  and  gravitational radiation  were  taken  into
account,        assuming       disrupted        magnetic       braking
\citep{schreiberetal10-1}. Also, tidal  evolution, circularization and
synchronization were considered.

For  those binary  systems in  which  the primary  component had  time
enough to  evolve to the WD  stage three situations can  be found. For
detached systems in  which no mass transfer  episodes occur whatsoever
we    adopted    the    initial-to-final    mass    relationship    of
\cite{cataetal2008} to  obtain the mass of  the WD. In those  cases in
which the mass transfer was  stable we employed the procedure detailed
in  \cite{hurleyetal02-1}, while  if the  mass transfer  was unstable,
i.e. if  the system underwent  a CE  phase, we followed  the procedure
detailed in Sect.~\ref{s-CEP}.

In  all  the  three   cases  previously  described  the  corresponding
evolutionary properties  of the resulting  WD must be  interpolated in
the appropriate cooling tracks. For  low-mass helium-core WDs (He WDs,
$M_{\rm WD} \la 0.5\, M_{\sun}$) we adopted the evolutionary tracks of
\cite{Serenelli_2001}.  For intermediate-mass  carbon-oxygen core  WDs
(C/O WDs,  $0.5 \la  M_{\rm WD}/M_{\sun}  \la 1.1$)  we used  the very
recent  cooling tracks  of  \cite{Renedo_10}, which  include the  most
up-to-date physical  inputs. Finally,  for the high-mass  end ($M_{\rm
WD} \ga  1.1 \, M_{\sun}$)  of the  WD mass distribution,  composed by
oxygen-neon core WDs  (O/Ne WDs), we adopted the  cooling sequences of
\cite{altaetal_2007}. All these cooling  tracks correspond to WDs with
pure hydrogen atmospheres.

\subsection{Evolution during the CE phase}
\label{s-CEP}

The evolution during the CE phase was computed following the treatment
of  \cite{hurleyetal02-1}. In  particular,  the  Roche-lobe radius  is
calculated according  to the  prescription of  \cite{eggleton83-1} and
during the overflow  episodes both rejuvenation and  ageing were taken
into account. The final separation of  a WD+MS pair after the CE phase
was obtained using the usual prescription:

\begin{equation}
\frac{a_{\rm f}}{a_{\rm i}}=\left( \frac{m_{\rm WD}}{M_{1}}\right) 
\left[ 1+\left( \frac{2}{\lambda\alpha_{\rm CE}r_{\rm L1}}\right) 
\left( \frac{M_{\rm env}}{M_{2}}\right)\right] ^{-1}
\label{CE}
\end{equation}

\noindent where $a_{\rm i}$ and $a_{\rm  f}$ are the initial and final
orbital separations, $M_{\rm env}$ is the  mass of the envelope of the
primary star at  the beginning of the CE phase  and $r_{\rm L1}=R_{\rm
L1}/a_{\rm i}$, where $R_{\rm L1}$ is the radius of the primary at the
onset of  mass transfer,  $\alpha_{\rm CE}$ is  the CE  efficiency and
$\lambda$ is  the binding energy  parameter. These two  parameters are
described in detail below.

The  CE   efficiency  parameter,  $\alpha_{\rm  CE}$,   describes  the
efficiency  of ejecting  the envelope,  namely, of  converting orbital
energy into kinetic energy to eject the envelope. We then have:

\begin{equation}
E_{\rm bind}=\alpha_{\rm CE}\Delta E_{\rm orb}
\end{equation}

\noindent where $E_{\rm  bind}$ is the binding energy  of the envelope
of  the primary,  usually  approximated by  the gravitational  energy,
i.e.:

\begin{equation}
E_{\rm bind}=-\int_{M_{\rm core}}^{M_{\rm donor}}\frac{GM(r)}{r}{\rm d}m,
\label{exact}
\end{equation}

\noindent generally rewritten in a more compact and suitable way as:

\begin{equation}
E_{\rm bind}=-\frac{GM_{\rm donor}M_{\rm env}}{\lambda R_1} 
\label{lambda}
\end{equation}

\noindent  where  $\lambda$ is  the  binding  energy parameter,  which
represents the ratio between the  approximate and the exact expression
of the binding energy. In passing,  we note that this approximation is
equivalent to  assume that the resulting  WD is a point  mass and that
the envelope  is a  shell of homogeneous  density located  at distance
$\lambda R_1$ from the core of the primary.

We recall here that \cite{han1995} introduced a parameter $\alpha_{\rm
th}$ to characterize the fraction of  the internal energy that is used
to expel  the CE. As in  \citet{Zorotovic_2010}, we will use  here the
notation $\alpha_{\rm  int}$ for this  parameter to emphasize  that it
includes  not only  the  thermal  energy but  also  the radiation  and
recombination energy. According to this, Eq.~(\ref{exact}) becomes:
 
\begin{equation}
E_{\rm bind}=\int_{M_{\rm core}}^{M_{\rm donor}}\left(-\frac{GM(r)}{r}+
\alpha_{\rm int}U_{\rm int}\right){\rm d}m 
\label{exactUint}
\end{equation}

\noindent One  can include the effects  of the internal energy  in the
binding  energy parameter  $\lambda$  by equating  Eqs.~(\ref{lambda})
and~(\ref{exactUint}). Thus, $\lambda$ clearly  depends on the mass of
the  donor, its  evolutive  stage  and the  fraction  of the  internal
energy,    $\alpha_{\rm   int}$,    available    for   ejecting    the
envelope. Except  for models in which  a fixed value of  $\lambda$ was
assumed, the values of $\lambda$ were computed using a subroutine from
the binary-star evolution (BSE) code from \cite{hurleyetal02-1}.

With  these  prescriptions  we  were   able  to  produce  a  synthetic
population of  WD+MS binaries. For  the rest  of this paper,  we focus
only on those systems that experienced a CE phase (PCEBs) and that are
still detached.

\subsection{Photometry}
\label{s-photom}

The  Monte  Carlo   simulator  described  so  far   does  not  provide
photometric magnitudes for the simulated  WD+MS PCEBs. In this section
we explain  how we obtain $ugriz$  SDSS magnitudes for the  two binary
components in an independent manner,  that are then combined to obtain
the magnitudes of the simulated sample of WD+MS PCEBs.

WD  Johnson-Cousins   $UBVRI$  magnitudes   were  obtained   from  the
evolutionary    tracks    detailed    in    the    previous    section
\citep{Serenelli_2001, Renedo_10, altaetal_2007}.  To transform to the
SDSS  $ugriz$ system  we  simply followed  the  procedure detailed  in
\cite{jordi_2006}. The photometry of  the companion stars was obtained
as follows. We first used the empirical spectral type-mass relation of
\citet{rebassa-mansergasetal07-1}  and obtained  the spectral  type of
the secondary stars  (note that the secondary star mass  is known from
the Mote Carlo simulator).  This relation is only defined for M-dwarfs
($M\la0.45\,M_{\sun}$),  however,  as  it  will  be  shown  later  (in
Sect.\,\ref{s-biasin}), WD+MS pairs  containing earlier type secondary
stars  are excluded  from the  simulated  sample as  a consequence  of
selection effects affecting the observed population of PCEBs. For each
spectral type we  then obtained average $u-g$, $g-r$,  $r-i$ and $i-z$
colors. These  were obtained fitting  a large sample of  SDSS M-dwarfs
\citep{westetal08-1}     to      the     M-dwarf      templates     of
\citet{rebassa-mansergasetal07-1}. Once $\sim25-30$  stars were fitted
for each spectral type, we  then calculated the above mentioned colors
using  the available  SDSS  un-reddened magnitudes  of the  considered
M-dwarfs  and averaged  them.  Our  average  colors are  in very  good
agreement  with those  of  \citet{westetal11-1} for  $g-r$, $r-i$  and
$i-z$.  For  $u-g$ this  exercise works  relatively well  for spectral
types M0$-$5, however  it becomes rather uncertain  for later spectral
types. To avoid this we searched for nearby late-type M-dwarfs ($>$M5)
in the sample of  \citet{bochanskietal11-1} with available un-reddened
magnitudes in SDSS and averaged  their $u-g$ colors. This dramatically
reduced the  uncertainties. Once the  average colors were  obtained we
used  the empirical  $M_r  - (r-i)$  and $M_r  -  (i-z)$ relations  of
\citet{bochanskithesis}  to obtain  $M_r$.   This,  together with  the
known  distance  from  the  Monte Carlo  simulator,  gives  $r$.   The
remaining $ugiz$  magnitudes were  easily calculated from  the average
colors.  We emphasize  that our procedure rests on  a purely empirical
basis,  thus avoiding  undesired biases  due to  the use  of synthetic
spectra,  which depend  mostly on  the surface  gravity and  effective
temperature, instead  of M-dwarf  template spectra,  which essentially
depend on the spectral type.

\begin{figure}
\centering
\includegraphics[width=\linewidth]{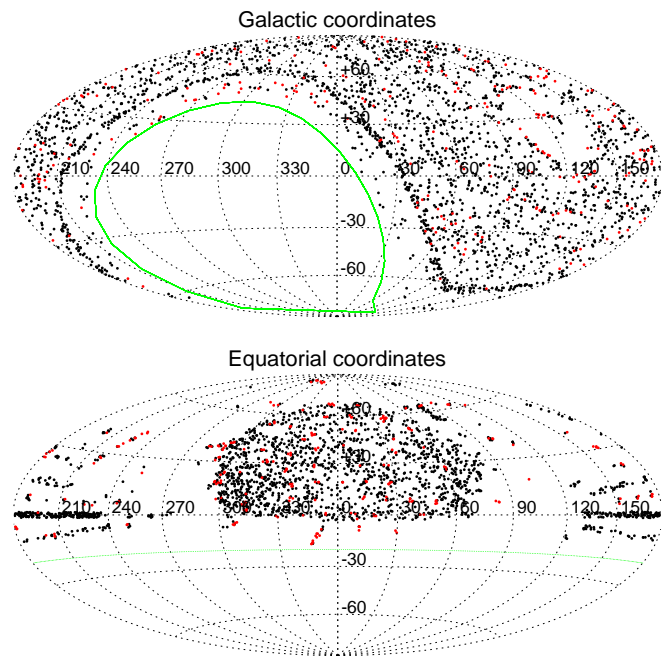}
\caption{Position of Legacy  (black) and SEGUE (red)  SDSS DR\,7 WD+MS
  binaries  in  Galactic  and   equatorial  coordinates.   Taken  from
  \citet{rebassa-mansergasetal12-1}.  See the  online  edition of  the
  journal for a color version of this figure.}
\label{f-plates}
\end{figure}

Once the  SDSS $ugriz$  magnitudes of the  two binary  components were
obtained, we added  the corresponding fluxes to  obtain the magnitudes
of the  simulated WD+MS PCEBs.  Finally, in order to  provide reliable
magnitudes  and colors  (see Sect.\,\ref{s-cuts})  Galactic extinction
was  incorporated using  the  model of  \cite{hakkila1997}, while  the
employed color correction was that of \cite{Schlegel}.


\section{Selection effects}
\label{bias}

So far we have described how we simulated the WD+MS PCEB population in
the Galaxy in the directions of the SDSS DR7 spectroscopic plates, and
how we  computed the SDSS  $ugriz$ magnitudes of the  entire simulated
sample. Given  that the  main purpose  of this paper  is to  perform a
detailed comparison  of the  simulated and  the observed  WD+MS binary
populations  in  the  SDSS  that  underwent a  CE  phase,  it  becomes
necessary to incorporate the observational selection effects in a very
realistic and detailed way. In this section we describe how we modeled
these selection biases.

\subsection{Color cuts}
\label{s-cuts}

Our first  step consisted in applying  a color filter. The  color cuts
allow  to   observationally  cull   WD+MS  binary  systems   from  the
spectroscopic      SDSS       DR7      WD+MS       binary      catalog
\citep{rebassa-mansergasetal12-1}.    From   this  sample,   we   only
considered   systems  observed   by  the   SDSS  Legacy   survey  (see
Fig.\,\ref{f-plates}), as WD+MS binaries  identified by SEGUE -- Sloan
Extension     for    Galactic     Understanding    and     Exploration
\citep{yannyetal09-1}   --  were   selected  following   a  completely
different algorithm \citep{rebassa-mansergasetal12-1}.  For magnitudes
within  the  range  $15<i<19.5$  the  color cuts  we  applied  to  the
synthetic    sample     were    the    following    --     see    also
\cite{rebassa-mansergasetal13-2}:

$$
\begin{array}{lll}
(u-g) & < & 0.93-0.27\times (g-r)-4.7\times(g-r)^2+\\
 & & +12.38\times(g-r)^3+3.08\times(g-r)^4-\\
 & & -22.19\times(g-r)^5+16.67\times(g-r)^6-\\
 & & -3.89\times(g-r)^7\nonumber
\end{array}
$$

$$
-0.5<(g-r)<1.7
$$

$$
-0.4<(r-i)<1.8\nonumber
$$

\noindent and

$$
\begin{array}{lllcl}
(g-r)&<&2\times(r-i)+0.38 & {\rm if} & -0.4<(r-i)\le0.1\\
(g-r)&<&0.5 & {\rm if} & 0.1<(r-i)\le0.3\\
(g-r)&<&4.5\times(r-i)-0.85 & {\rm if} & 0.3<(r-i)\le0.5\\
(g-r)&<&0.25\times(r-i)+1.3 & {\rm if} & 0.5<(r-i)\le1.8\nonumber
\end{array}
$$

\subsection{Spectroscopic completeness}
\label{s-spectro}

The  main science  driver of  the SDSS  Legacy survey  was to  acquire
spectroscopy    for    magnitude-limited     samples    of    galaxies
\citep{straussetal02-1} and quasars \citep{richardsetal02-1}.  Because
of their  composite nature,  WD+MS binaries form  a ``bridge''  in the
color space  that connects the  WD locus  with that of  low-mass stars
\citep{smolcicetal04-1}. The blue end  of the bridge, characterized by
WD+MS  binaries with  hot WDs  and/or late  type companions,  strongly
overlaps  with  the   color  locus  of  quasars,   and  was  therefore
intensively targeted for  spectroscopy by the SDSS  Legacy Survey.  In
contrast, the  red end of  the bridge  is dominated by  WD+MS binaries
containing cool WDs,  and excluded from the quasar  program. Thus, the
next step in producing realistic simulations of the PCEB population is
to  apply  a spectroscopic  completeness  correction  that takes  into
account the  probability of  a given  simulated PCEB  with appropriate
colors to be spectroscopically observed by the SDSS Legacy survey.

To  estimate  this probability  we  proceeded  as follows.   We  first
calculated  the  spectroscopic  completeness   of  each  WD+MS  binary
observed by the SDSS DR\,7 Legacy  Survey.  It is important to keep in
mind  that these  observed WD+MS  binaries include  wide systems  that
never interacted during their evolution and PCEBs, and that only PCEBs
are considered in  the numerical sample.  Strictly  speaking we should
consider  then only  those  observed WD+MS  binaries  that are  PCEBs.
However, the number  of identified PCEBs is just $\sim$10  per cent of
the       entire        SDSS       WD+MS        binary       catalogue
\citep{rebassa-mansergasetal13-2}, and we know that about one third of
the   total   number   of   WD+MS    binaries   should   be   a   PCEB
\citep{schreiberetal10-1}.   Besides, there  is no  reason to  believe
that the spectroscopic completeness will vary from wide to close WD+MS
binaries.  In order to avoid low number statistics in our calculations
we thus  decided to use the  entire observed sample, i.e.   wide WD+MS
plus PCEBs.  We  did exclude however WD+MS binaries  that are resolved
in their SDSS  images, as these are associated  to large uncertainties
in their photometric magnitudes.  The resulting sample contains 1\,645
systems.

We obtained  the $u-g$, $g-r$, $r-i$  and $i-z$ colors of  each of our
1\,645  observed WD+MS  binaries,  and defined  a four-dimension  (one
dimension per color)  sphere of 0.2 color radius around  each of them.
Within   each   sphere  we   calculated   via   DR\,7  {\tt   casjobs}
\citep{li+thakar08-1}  the   number  of   point  sources   with  clean
photometry ($N_\mathrm{phot}$) as well  as the number of spectroscopic
sources  ($N_\mathrm{spec}$).  This  search  was  restricted to  those
systems  fulfilling the  color cuts  given in  Sect.~\ref{s-cuts}. The
choice of  a sphere  radius of 0.2  ensures that  $N_\mathrm{spec}$ is
larger than 15  in each case.  The spectroscopic  completeness of each
of    the    observed   WD+MS    systems    is    simply   given    by
$N_\mathrm{spec}$/$N_\mathrm{phot}$.  The probability  for a simulated
PCEB to  be observed  spectroscopically by the  Legacy survey  of SDSS
finally corresponds to the  spectroscopic completeness of the observed
WD+MS  binary with  the most  similar colors,  i.e. the  closest color
distance (as defined  by the four colors) between  the simulated WD+MS
binary and  the observed systems.  After applying the  color selection
filter, the synthetic binaries  occupy color regions densely populated
by  the  observed  WD+MS  binaries.  We find  that,  on  average,  the
four-dimensional  color  distance  from  one synthetic  WD+MS  to  the
nearest observed target  is 0.09, a rather  reasonable value, although
this distance can be  in some cases as small as  0.01, whereas only in
$\sim 4\%$ of the cases it is larger than 0.2.

\subsection{Intrinsic WD+MS binary bias}
\label{s-biasin}

It is  expected that a certain  fraction of the simulated  WD+MS PCEBs
should contain primary  or secondary stars that  would be undetectable
in the spectrum if observed spectroscopically by the SDSS. This is the
case when one of the  stellar components is considerably brighter than
the other and overshines the companion. For late-type secondary stars,
this implies an upper limit on  the WD effective temperature, at which
we  would be  able  to discern  the companion  in  the SDSS  spectrum.
Conversely, the detection of WDs next to early-type companions results
in a lower  limit on the WD effective temperature.   In addition, SDSS
spectra  of farther  objects are  associated to  lower signal-to-noise
ratio. Our  observed sample of WD+MS  PCEBs is partially based  on the
visual identification of  both binary components in  the SDSS spectrum
\citep{rebassa-mansergasetal10-1}, and  consequently objects  with low
signal-to-noise ratio may have not passed the identification criteria.
This implies an upper limit in  the distance of WD+MS binaries.  These
two effects need  to be taken into account in  our simulated sample of
WD+MS PCEBs.

In order to evaluate the above described selection effects we followed
the approach adopted by \citet{rebassa-mansergasetal11-1} and used the
WD  atmosphere  models  of  \citet{koesteretal05-1}  and  the  M-dwarf
templates  of  \citet{rebassa-mansergasetal07-1} to  obtain  synthetic
composite WD+MS binary spectra in  the wavelength range and resolution
provided by  typical SDSS  spectra for  a wide  range of  WD effective
temperatures ($T_\mathrm{eff}$ ranging from 6\,000 to 100\,000~K in 37
steps  nearly  equidistant  in   $\log  T_\mathrm{eff}$)  and  surface
gravities  (covering  from $\log  g=6.5$  to  9.5  in steps  of  0.5),
spectral type  of the companions  (M0$-$9, in steps of  one subclass),
and  distances (from  50  to  1\,700~pc in  steps  of  50~pc). To  the
complete  set  of  synthetic  composite spectra  we  added  artificial
Gaussian noise  varying according to the  distance used. Specifically,
the noise  level introduced  to the  composite spectra  reproduces the
signal-to-noise ratio that  the observed WD+MS binary  spectra have at
the considered distance.

Once the  synthetic spectra were  obtained, we subjected  the complete
sample to  the identification criteria  defined for real  WD+MS binary
spectra in  SDSS, namely  a visual  inspection of  the spectra,  and a
search for blue and red excess  in those spectra dominated by the flux
of  the  secondary  star  and   WD  components,  respectively  --  see
\cite{rebassa-mansergasetal10-1}   for   details.   In   addition   we
calculated  the  $ugriz$ magnitudes  from  the  synthetic spectra  and
excluded  all   systems  exceeding  the  magnitude   limits  given  in
Sect.\,\ref{s-cuts}. From  the resulting sample we  then evaluated the
WD effective  temperature and distance  limits that were  then applied
accordingly to  the sample of  WD+MS binaries obtained from  the Monte
Carlo simulator.

\subsection{PCEB orbital period filter}
\label{s-orbital}

\begin{figure}[t]
\vspace{8cm} 
\includegraphics{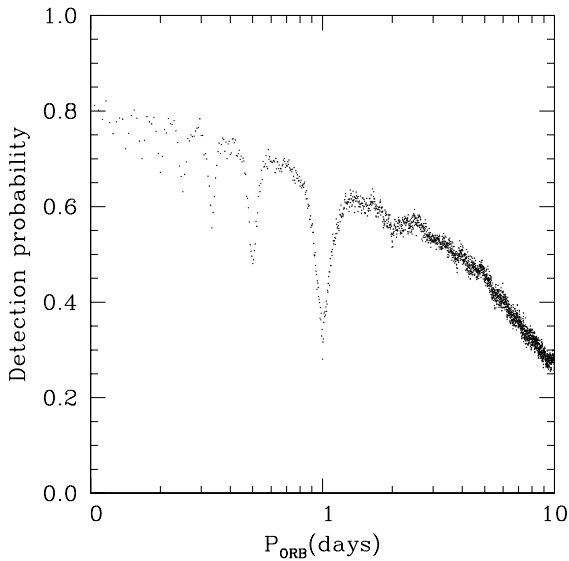}
\caption{Detection probability of a PCEB  as a function of the orbital
  period.}
\label{f:period}
\end{figure}

\begin{figure*}[t]
\vspace{11cm}
\includegraphics{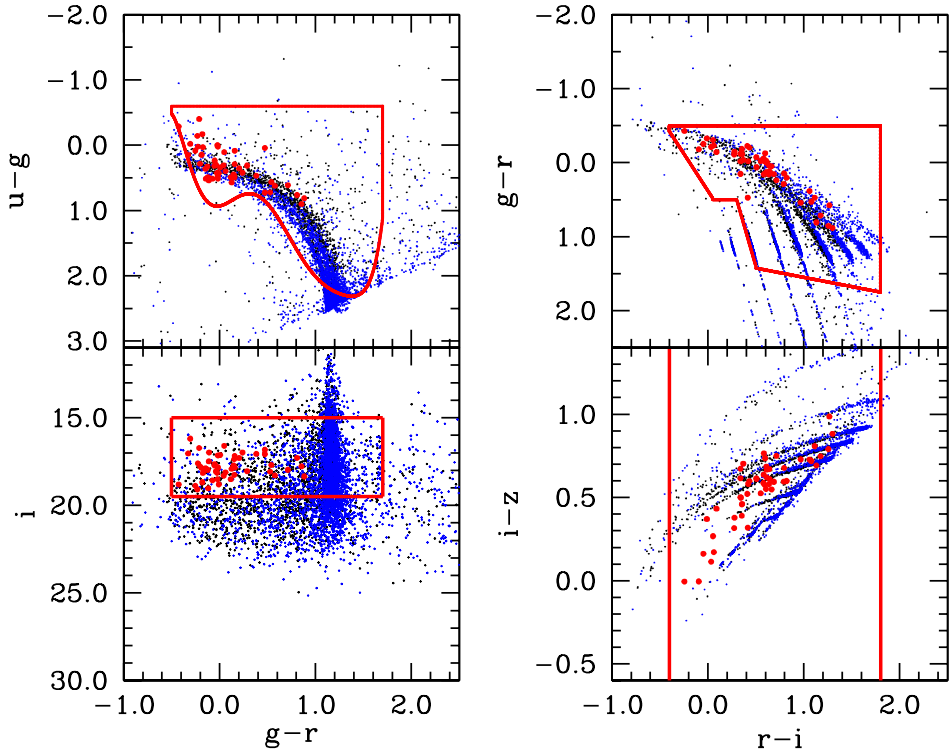} 
\caption{Color-color  diagram of  the synthetic  WD+MS PCEBs  obtained
  using   our  Monte   Carlo  simulator   when  our   reference  model
  ($\alpha_{\rm  CE}=1.0$, $\lambda=0.5$,  and $n(q)=1$)  is employed.
  Systems containing  He WDs are  represented using black  dots, while
  blue dots correspond to systems with  C/O or O/Ne WDs.  The observed
  WD+MS  PCEB  systems  are  displayed  using  red  dots.   The  color
  selection     criteria     are     shown     using     red     lines
  (section\,\ref{s-cuts}). See the online version of the journal for a
  color version of this figure.}
\label{f:cdm}
\end{figure*}

Finally,  we filtered  our  simulated binary  systems  according to  a
period  efficiency   function,  which  measures  the   probability  of
identifying  a  PCEB  among  the WD+MS  SDSS  sample.   The  detection
probability    function    \citep{nebotetal11-1}     is    shown    in
Fig.~\ref{f:period}.  As  can be  seen, the  probability of  finding a
binary system decreases for increasing  periods, and drops rapidly for
those systems with  period larger than 3 days. For  orbital periods of
one day or multiples of one  day the probability for sampling the same
orbital phase increases, which translates  in a decrease of the period
efficiency function.


\section{The observed sample}
\label{s-sample}

The sample of  binary systems that we use for  comparison consisted on
53 WD+MS PCEBs  from the SDSS DR7 catalogue with  known periods -- see
\cite{rebassa-mansergasetal12-1},       \cite{nebotetal11-1}       and
\cite{Zorotovic_2010},   and  references   therein.   As   we  already
mentioned,  SEGUE systems  have been  excluded. The  periods are  well
determined, and  therefore the  distribution of  periods is  useful to
compare  with  the  period  distribution obtained  for  the  simulated
systems. To compare with our models, we are also interested in knowing
the core  composition of the WD  in the observed systems,  to estimate
the fraction  of systems  containing He  WDs, and  also the  number of
systems containing more  massive O/Ne WDs. To do this  we proceeded as
follows.   If the  mass of  a WD  is smaller  than $0.5\,M_{\sun}$  we
assumed that it has  a He core. {\bf Conversely}, if the  mass of a WD
is larger than $0.5\,M_{\sun}$ but  smaller than $1.1\,M_{\sun}$ a C/O
core  was adopted.   Finally, if  the mass  of the  WD is  larger than
$1.1\,M_{\sun}$ an O/Ne  core was adopted.  For 49 of  the 53 PCEBs in
the sample, it has been possible to determine the mass of the WD using
the  method  described  by \citet{rebassa-mansergasetal07-1}.   As  in
\citet{Zorotovic_2010}, in  order to  determine their  compositions we
decided  to exclude  systems  with WD  temperatures  below 12\,000  K,
because the spectral  fitting methods are not reliable  for cooler WDs
and  therefore their  masses can  not be  trusted.  This  implies that
reliable WD masses can be obtained for  40 of the 53 systems that form
our observed sample,  of which 14 have a  He WD, 23 a C/O WD  and 2 an
O/Ne  WDs.   There   is  also  one  system   with  $M_{\mathrm{WD}}  =
0.5\,M_{\sun}$  for which  we  can  not decide  which  type  of WD  it
is. This  corresponds to a  fraction of $36\pm 8\%$  of He WDs  in the
sample,  where   we  have   assumed  binomial  errors.    This  issue,
nevertheless,    will    be    discussed    in    more    detail    in
Sect.~\ref{s-periods}.


\begin{figure*}[t]
\vspace{11cm}
\includegraphics{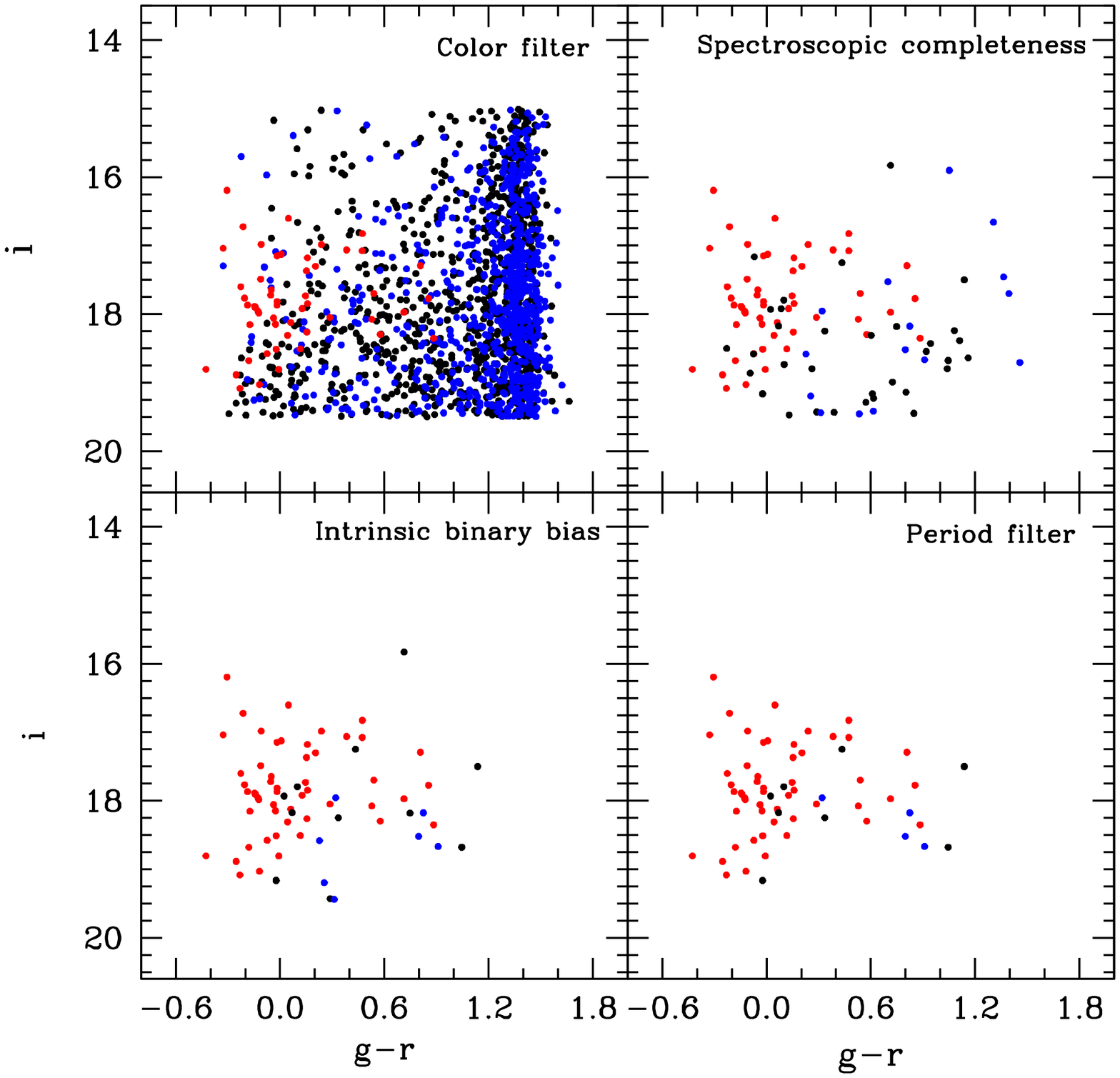} 
\caption{Color-magnitude diagram of the synthetic WD+MS PCEBs obtained
  using our Monte Carlo simulator  (blue and black dots) compared with
  the  observed systems  (red  symbols) after  applying the  different
  filters explained in the text to our reference model. Colors are the
  same as in  Fig.~\ref{f:cdm}. See the online version  of the journal
  for a color version of this figure.}
\label{f:mc}
\end{figure*}

\section{Results}
\label{s-results}

We computed  a large  number ($\sim 500$)  of Monte  Carlo simulations
covering  a wide  range  of  values of  the  CE efficiency  parameter,
$0.0\leq \alpha_{\rm  CE}\leq 1.0$, and  the fraction of  the internal
energy available  to eject the  CE, $0.0\la \alpha_{\rm  int}\la 0.3$,
which can result in very large values of the binding energy parameter,
$\lambda$.   We  also performed  simulations  in  which $\lambda$  was
computed  including the  contribution  of different  fractions of  the
internal energy, $\alpha_{\rm  int}$. All this was done  for the three
IMRDs,  $n(q)$, previously  mentioned in  Sect.\, \ref{general}.   For
each  of  our  models  we   generated  $10$  independent  Monte  Carlo
simulations (with different initial seeds) and for each of these Monte
Carlo realizations, we  increased the number of  simulated Monte Carlo
realizations to  $10^4$ using bootstrap techniques.   Specifically, in
all  our  calculations we  used  the  resampling method  described  in
\cite{Chernick_2007}.   The method  consists  in generating  resamples
with  a  probability equal  to  that  of  the original  sample.   Each
resample, also  called a  ``bootstrap sample" or  ``replication", must
have the same size (number of  elements) as the original sample.  This
is   the   reason  why   this   method   is  named   resampling   with
replacement.  Due to  the fact  that  resampling can  be done  without
adopting any particular assumption  about the probability distribution
of the population,  this technique can be used not  only to derive the
sample distribution-free  values of  interest, but also  for assessing
the precision  and variability  of sample statistics.  In this  way we
were  able to  streamline  the Monte  Carlo  calculations, with  large
savings of  computer time. Moreover,  using this procedure  we ensured
convergence in  all the  final values of  the relevant  quantities. In
what follows we  describe the model predictions and  compare them with
the observations.  Given that the  parameter space of CE  evolution is
very large, we show in this  paper only those results which imply some
relevant differences between the corresponding models.

\subsection{Color-color space}

We first investigate  whether the simulated PCEB  population is placed
in the same regions in the color-color space as the observed PCEBs. To
that end,  and for  the sake  of definiteness,  we define  a reference
model for  which we  considered $\alpha_{\rm  CE}=1.0$, $\lambda=0.5$,
and a  flat IMRD, $n(q)=1$.  This  choice of parameters should  not be
considered as a representative case, and  we use it just to illustrate
the effects of the different filters applied to the simulated samples.
Moreover,  we adopted  this  model because  it  represents an  extreme
(albeit frequently employed)  case among the many  possible choices of
the free parameters of  common envelope evolution.  Figure~\ref{f:cdm}
shows an example of the color-color diagram of present-day WD+MS PCEBs
obtained in a typical Monte Carlo realization for our reference model.
Systems  which underwent  the CE  phase  before He  ignition (case  B)
contain  He WDs  and are  represented  by black  dots.  Systems  which
underwent the CE episode after He  ignition during the early AGB (case
C) or  during the thermally pulsing  AGB phase (TPAGB) contain  C/O or
O/Ne WDs  and are displayed using  blue dots.  Red dots  correspond to
the observed WD+MS  PCEBs.  The different color cuts  discussed in the
Sect.~\ref{s-cuts}  are represented  by red  lines.  A  quick look  at
Fig.~\ref{f:cdm} reveals that our  simulations recover fairly well the
observed  population  of  WD+MS  PCEBs in  the  different  color-color
diagrams, and  that our  synthetic WD+MS PCEBs  overlap with  the real
ones.   Moreover,  our simulated  population  lies  within the  region
allowed by the different color cuts.  However, as expected, the entire
simulated Galactic population  of PCEBs occupies a  region larger than
the   observational   one,   especially   in  the   $i$   vs.    $g-r$
color-magnitude diagram.  Finally,  we note as well  that the discrete
blue  tracks come  from the  fact that  we are  mapping MS  stars onto
discrete spectral types.

\subsection{The effects of biases and selection criteria}
\label{s-biases}

\begin{table*}[t]
\begin{center}
\caption{Total number and percentage of simulated WD+MS binary systems
  obtained  after applying  the  successive  selection criteria.   For
  model  1  we  adopted   $\alpha_{\rm  CE}=1.0$,  $\lambda=0.5$,  and
  $n(q)=1$,  for  model  2  the   set  of  theoretical  parameters  is
  $\alpha_{\rm CE}=1.0$, $\lambda=0.5$, and $n(q)=q^{-1}$, whereas for
  model 3  we employed  $\alpha_{\rm CE}=0.3$, $\lambda$  was computed
  assuming $\alpha_{\rm int}=0.2$, and $n(q)=q^{-1}$.}
\label{t:filters}
\centering
\begin{tabular}{lrrrrrr}
\hline
\hline
\multicolumn{1}{c}{\ } &
\multicolumn{5}{c}{Model 1}\\
\cline{2-6}
 & He & C/O - O/Ne & Total & Filtered (\%) & Cumulative (\%)\\
\cline{2-6} 
Unfiltered sample & 8\,344 (36\%) & 14\,834 (64\%) & 23\,178 & --- & 100\\
Color cuts & 980 (57\%) & 740 (43\%) & 1\,720 & 7.42 & 7.42 \\
Spectroscopic completeness & 35 (70\%) & 15 (30\%) & 50 & 2.91 & 0.21\\
Intrinsic binary bias & 13 (65\%) & 7 (35\%) & 20 & 40.00 & 0.86\\
Period filter & 8 (67\%) & 4 (33\%) & 12 & 60.00 & 0.05\\
\hline
\multicolumn{1}{c}{\ } &
\multicolumn{5}{c}{Model 2}\\
\cline{2-6}
 & He & C/O - O/Ne & Total & Filtered (\%) & Cumulative (\%)\\ 
\cline{2-6} 
Unfiltered sample & 12\,499 (30\%) & 28\,890 (70\%) & 41\,389 & --- & 100\\
Color cuts & 1\,478 (52\%) & 1\,365 (48\%) & 2\,843 & 6.87 & 6.87\\
Spectroscopic completeness & 66 (62\%) & 41 (38\%) & 107 & 3.76 & 0.26\\
Intrinsic binary bias & 22 (58\%) & 16 (42\%) & 38 & 35.51 & 0.09\\
Period filter & 14 (61\%) & 9 (39\%) & 23 & 60.52 & 0.06\\
\hline
\multicolumn{1}{c}{\ } &
\multicolumn{5}{c}{Model 3}\\
\cline{2-6}
 & He & C/O - O/Ne & Total & Filtered (\%) & Cumulative (\%)\\ 
\cline{2-6} 
Unfiltered sample & 17\,674 (25\%) & 53\,023 (75\%) & 70\,697 & --- & 100 \\
Color cuts & 2\,596 (47\%) & 2\,927 (53\%) & 5\,523 & 7.81 & 7.81\\
Spectroscopic completeness & 126 (56\%) & 99 (44\%) & 225 & 4.03 & 0.32\\
Intrinsic binary bias & 40 (55\%) & 33 (45\%) & 73 & 32.44 & 0.10\\
Period filter & 28 (65\%) & 15 (35\%) & 43 & 58.90 & 0.06\\
\hline
\end{tabular}
\end{center}
\end{table*}

The  effect  of  each  filter   over  the  simulated  WD+MS  PCEBs  is
illustrated in  Fig.~\ref{f:mc} for our  reference model in  the $g-r$
versus $i$ color-magnitude diagram.  Each panel represents the systems
that survive after consecutively applying  the filter indicated on it.
We show the  effect of the color selection  filter (upper-left panel),
the  result   of  applying   the  spectroscopic   completeness  filter
(upper-right panel)  to the previous  sample, the effect of  using the
intrinsic  binary  bias filter  (lower-left  panel),  and finally  the
result after  using the period  filter (lower-right panel). As  can be
seen, the different  filters applied to the  original synthetic sample
(black and blue  dots) severely reduce the total  number of observable
objects, which is  consistent (within an order of  magnitude) with the
observed sample (red  symbols).  As can be seen, the  final sample for
this specific Monte  Carlo simulation shows a poor  agreement with the
observed one. In particular, for this specific simulation the observed
binaries occupy  a region  that is  systematically bluer  and brighter
than  that  of   the  synthetic  sample.   The  reason   for  this  is
twofold.  The   first  reason  is  the   otherwise  natural  intrinsic
dispersion of  any Monte Carlo  simulation. Given the small  number of
synthetic  binaries  surviving  the  different cuts,  and  the  scarce
observational data, these effects can become prominent in a particular
Monte Carlo  sample. However,  the most important  reason is  that, as
already mentioned, the set of  theoretical parameters adopted for this
specific model --  namely the choice of  $\alpha_{\rm CE}$, $\lambda$,
and $n(q)$  -- is  clearly excluded  by the  observations and  we only
adopted it  for illustrative  purposes, given  that this  is a  set of
parameters  that  is  frequently  employed  in  the  literature.  More
elaborated models,  which fit better  the observational data,  will be
discussed below, in Sects.~\ref{sec:KS}, and \ref{sec:p-md}.

In  order  to quantitatively  analyze  the  effects of  the  different
selection criteria on the entire  population of simulated WD+MS PCEBs,
we show in  table~\ref{t:filters} the total number  and percentage (in
parentheses)  of WD+MS  PCEBs initially  simulated and  obtained after
applying consecutively the selection criteria and observational biases
described in Sects.~\ref{s-cuts} to~\ref{s-orbital}.   We also list in
the last column  of this table the cumulative percentage  of the WD+MS
population obtained  after applying  the selection  cuts. We  show the
results for  three representative  models.  Model  1 is  our reference
model, previously  described.  In  model 2  we also  used $\alpha_{\rm
CE}=1.0$ and  $\lambda=0.5$, but  we adopted $n(q)\propto  q^{-1}$, to
illustrate the effects  of the IMRD.  Finally, for model  3 we adopted
$\alpha_{\rm CE}=0.3$,  and $n(q)\propto q^{-1}$, while  $\lambda$ was
computed  for  every  binary  assuming  $\alpha_{\rm  int}=0.2$.   The
unfiltered  samples, which  correspond to  the total  number of  WD+MS
PCEBs in the SDSS DR7 fields irrespective of their apparent magnitude,
are sufficiently large  in all three cases, and allow  us to study the
effects  of the  succesive filters.   As  can be  seen, the  selection
criteria produce a dramatic decrease  of the total number of simulated
WD+MS PCEBs, independently  of the adopted model.   In particular, the
final  simulated population  is smaller  than $0.1\%$  of the  initial
sample for all three models -- see the last column of this table.  The
most  restrictive  selection  criteria  are the  color  cuts  and  the
spectroscopic completeness filter.  Only $\sim  7\%$ of the objects in
the input sample  pass the cuts in colors and  magnitude for all three
models, while  the spectroscopic completeness filter  eliminates $\sim
97\%$  of those  that survive  the first  filter.  If  only these  two
filters  are applied  the total  population of  potentially observable
systems decreases  drastically down  to $0.2-0.3\%$ of  the unfiltered
sample.  This behavior can be  easily explained.  First, the SDSS only
covered  $15<i<19.5$  and  most  WD+MS  binaries  in  our  Galaxy  are
obviously fainter.  Second, the SDSS  was primarily designed to detect
galaxies  and quasars  and thus  the  probability for  a WD+MS  binary
system  to be  spectroscopically detected  by the  SDSS is  relatively
small,   specially   for   WD+MS    binaries   containing   cool   WDs
\citep{rebassa-mansergasetal13-2}.  The  remaining filters,  i.e.  the
intrinsic binary bias filter and the period filter, further reduce the
size  of the  sample of  simulated PCEBs  systems. In  particular, the
intrinsic binary bias  filter reduces the number  of systems surviving
the spectroscopic  completeness filter to about  $30-40\%$, whilst the
period  filter  reduces  the  sample   of  systems  that  survive  the
spectroscopic   completeness  filter   to  $\sim\,60\%$.    Thus,  the
selection criteria play a crucial role since only $\sim 0.05\%$ of the
simulated binary systems survive the successive filters.

The final number of WD+MS PCEBs predicted to be identified by the SDSS
is in  reasonable agreement with  the observed number of  systems (see
table~\ref{t:filters}).    This  indicates   that  both   our  initial
assumptions as  well as the  computation of the selection  effects and
biases are  likely good  representations of  reality.  However,  it is
important  to  realize that  the  number  of predicted  PCEBs  depends
somewhat  on the  adopted values  of $\alpha_{\rm  CE}$ and  $\lambda$
during the CE  phase. We obtain the best agreement  (i.e.  the largest
number  of  predicted  systems)  assuming a  variable  binding  energy
parameter and a small CE efficiency, namely for model 3.

Interestingly, the  selection criteria  employed to select  the sample
introduce  an unexpected  bias  in the  observed  population of  WD+MS
PCEBs, as the  fraction of systems containing He WDs  that are finally
culled from the total population increases independently of the model,
from $\sim 25-35\%$ to $\sim  60-70\%$. This implies that the observed
population of WD+MS  PCEBs is severely biased as a  consequence of the
selection  criteria  employed  to  cull   it,  and  that  WD+MS  PCEBs
containing  a  He  WD  are   over-represented  in  the  final  sample,
independently of the adopted model, due to the observational selection
effects.

\subsection{The role of the enhanced mass-loss parameter}

It  has been  suggested \citep{te1988}  that the  presence of  a close
companion could enhance mass loss during the red giant phase. As shown
in Eq.~(1)  the mass-loss  tidal enhancement  depends on  a parameter,
$B_{\rm W}$, which at present is unknown. To evaluate the influence of
this  parameter on  the resulting  population of  WD+MS PCEBs,  and to
better constrain the value of this enhancement parameter, we performed
an additional  set of simulations  in which we adopted  several values
for $B_{\rm W}$, ranging from 0  (no tidal enhancement) to $10^3$. The
results of  such simulations are presented  in Table~\ref{t:enhanced},
where we  show the percentages  of He and C/O  (or O/Ne) WDs  in WD+MS
PCEBs  for several  values  of  $B_{\rm W}$,  after  applying all  the
selection  effects  to  the   three  models  previously  described  in
Sect.~\ref{s-biases}. These  percentages are computed as  the ensemble
average  of a  sufficiently  large number  of  individual Monte  Carlo
realizations,  for which  we also  compute the  corresponding standard
deviations. Both are listed in Table~\ref{t:enhanced}. In general, the
percentage of He WDs increases as  $\log B_{\rm W}$ increases.  The He
WD fraction  increases because  as $B_{\rm W}$  is increased  the mass
losses are  larger. Increased mass  loss leads  to an increase  of the
orbital separation  and increases the  mass ratio $q$. This  can cause
the systems  not to  fill their  Roche-lobe, to end  up with  a longer
orbital period, or to stable mass transfer instead of evolving through
a CE  episode.  These effects  are strongest for systems  that without
increased mass  loss would fill their  Roche-lobe on the AGB  as these
systems  evolve   through  the   entire  sub-giant  and   first  giant
branch. Thus, increased  mass loss leads to a reduced  fraction of C/O
and O/Ne white dwarfs in PCEBs  while the number of PCEBs that contain
He-core WDs remains approximately constant.  We stress that even for a
small value of the enhancement parameter,  the percentage of He WDs is
somewhat large, at  odds with the observational data set  we are using
to compare,  for which  the fraction of  He WDs is  $\sim 40\%  $ (see
Sect.~\ref{s-sample}). Consequently, small values  of $B_{\rm W}$ seem
to be more compatible with the  observational data. For this reason in
the simulations  described in what  follows we adopted  $B_{\rm W}=0$,
which is a conservative choice.

\begin{table}[t]
\begin{center}
\caption{Enhanced  mass-loss parameter  and percentage  of PCEBs  with
  different types of WDs.}
\label{t:enhanced}
\begin{tabular}{lcccc}
\hline
\hline
\multicolumn{1}{c}{\ } &
\multicolumn{4}{c}{Model 1}\\
\cline{2-5}
$B_{\rm W}$ & 0 & 10 & $10^{2}$ & $10^{3}$ \\
\cline{2-5} 
He (\%) & $67\pm 12$ & $72\pm 8$ & $76\pm 8$ & $77\pm 8$ \\
C/O - O/Ne (\%) & $33\pm 12$ & $28\pm 8$ & $24\pm 8$ & $23\pm 8$ \\
\hline
\multicolumn{1}{c}{\ } &
\multicolumn{4}{c}{Model 2}\\
\cline{2-5}
$B_{\rm W}$ & 0 & 10 & $10^{2}$ & $10^{3}$ \\
\cline{2-5} 
He (\%) & $61\pm 10$ & $62\pm 7$ & $65\pm 7$ & $76\pm 7$ \\
C/O - O/Ne (\%) & $39\pm 10$ & $38\pm 7$ & $35\pm 7$ & $24\pm 7$ \\
\hline
\multicolumn{1}{c}{\ } &
\multicolumn{4}{c}{Model 3}\\
\cline{2-5}
$B_{\rm W}$ & 0 & 10 & $10^{2}$ & $10^{3}$ \\
\cline{2-5} 
He (\%) & $61 \pm 7$ & $65\pm 11$ & $68\pm 7$ & $75\pm 8$ \\
C/O - O/Ne (\%) & $39 \pm 7$ & $35\pm 11$ & $32\pm 7$ & $25\pm 8$ \\
\hline
\end{tabular}
\end{center}
\end{table}

\subsection{The effects of the internal energy}

\begin{figure*}[t]
\vspace{10.0cm}
\includegraphics{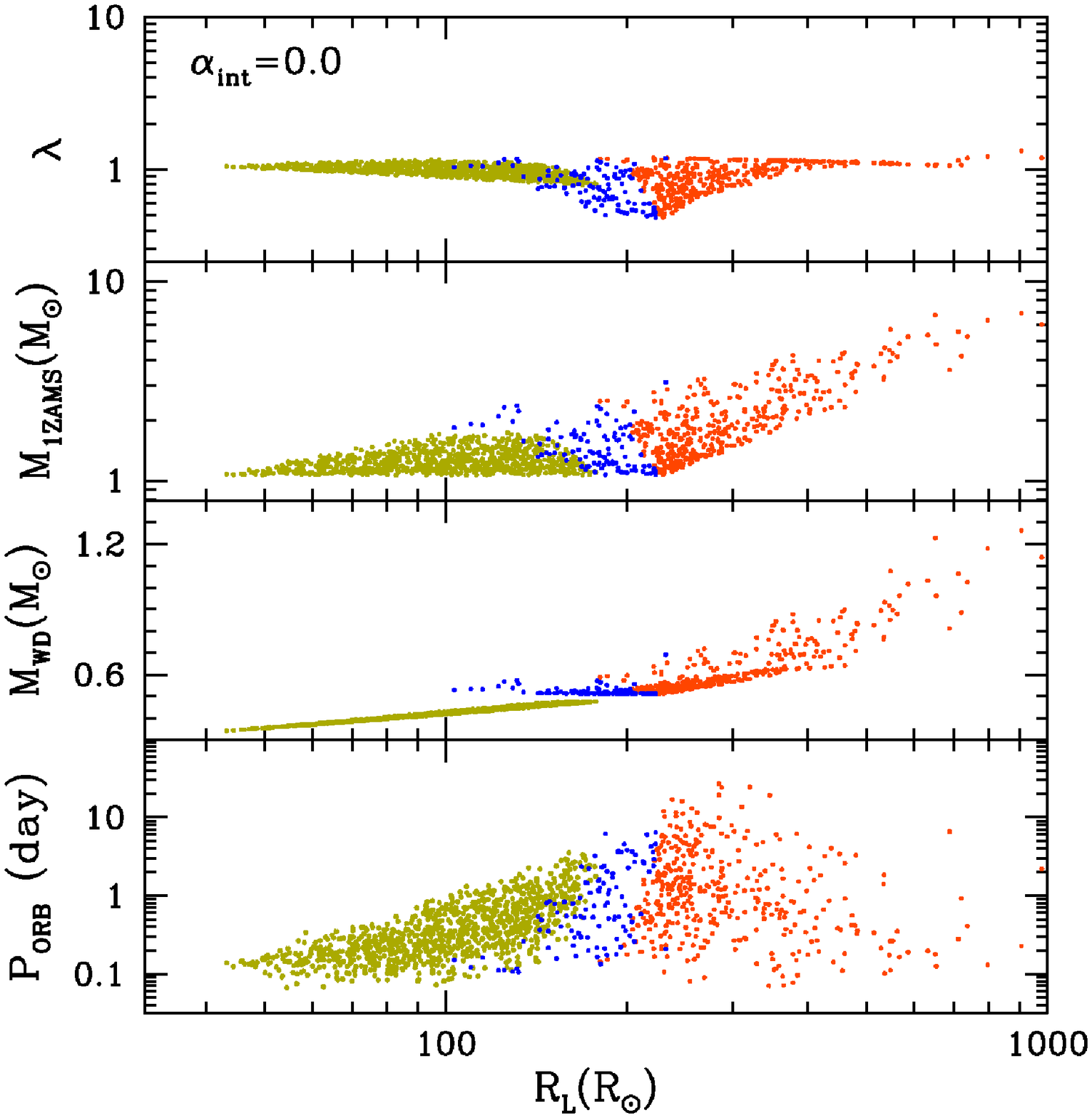}
\includegraphics{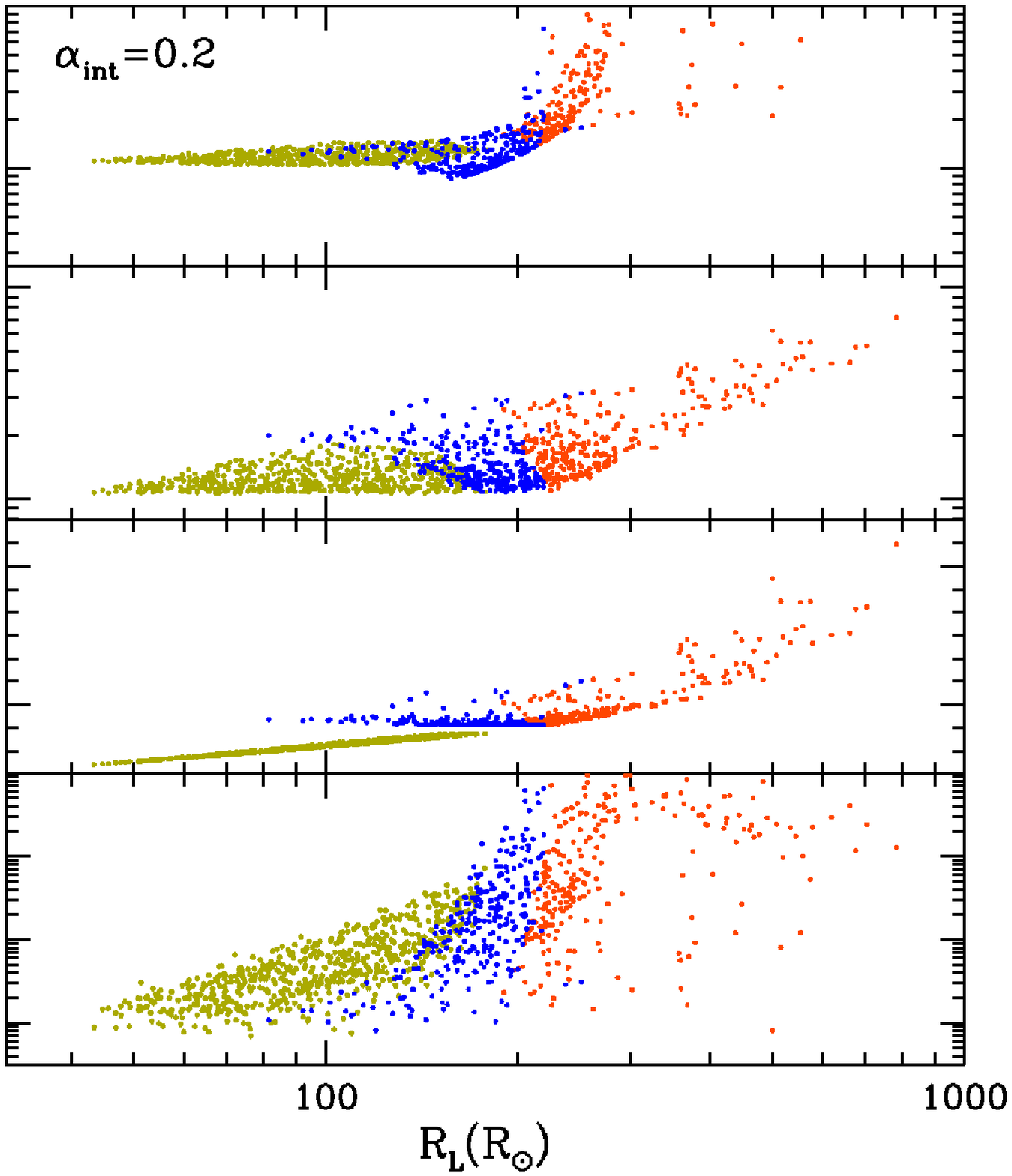}
\caption{From top  to bottom:  binding energy parameter,  primary ZAMS
  mass, WD  mass and orbital  period as  a function of  the Roche-lobe
  radius of the primary, as given in \cite{eggleton83-1}.  The case B,
  case C and TPAGB case CE  episodes are represented using green, blue
  and red dots, respectively.  The two panels show the results for two
  models in which  $\alpha_{\rm CE}=0.3$ and $n(q)=1$  but without and
  with a  fraction of  the internal energy  contributing to  expel the
  envelope:  $\alpha_{\rm  int}=0.0$  (left panels)  and  $\alpha_{\rm
  int}=0.2$ (right panels).  See the electronic version of the journal
  for a color version of this figure.}
\label{f:lambdas}
\end{figure*} 

\begin{table*}[t]
\begin{center}
\caption{Percentage of systems  with He WDs and KS test  of the period
  distribution for six representative models with $\lambda=0.5$.
\label{t:kstest:fixed}}
\footnotesize
\begin{tabular}{lcccccc}
\hline
\hline 
& \multicolumn{2}{c}{$n(q)\propto q^{-1}$} & \multicolumn{2}{c}{$n(q)=1$} & \multicolumn{2}{c}{$n(q)\propto q$} \\
\hline
$\alpha_{\rm CE}$ &1.0 & 0.25 & 1.0 & 0.25 & 1.0 & 0.25 \\
He(\%) & $67\pm 12$ & $47\pm 15$ & $61\pm 10$ & $47 \pm 12$ & $70\pm 10$ & $45\pm 12$ \\
KS & $0.46\pm 0.31$ & $0.53\pm 0.31$ & $0.54\pm 0.30$ & $0.56\pm 0.29$ & $0.58\pm 0.29$ & $ 0.54\pm 0.29$ \\
\hline
\end{tabular}
\end{center}
\end{table*}

Since more  than a  decade we  know that  assuming a  constant binding
energy  parameter  $\lambda$  is  probably not  a  good  approximation
\citep{dewi2000}.  Instead $\lambda$ depends on  the mass of the donor
star  and   the  evolutionary  stage.    We  explore  this   issue  in
Fig.~\ref{f:lambdas}   where  we   show  from   top  to   bottom:  the
distributions  of the  binding energy  parameter ($\lambda$),  primary
ZAMS masses, WD masses and periods, as a function of the radius of the
primary  just  prior  to  the  CE episode,  i.e.   of  its  Roche-lobe
radius. We compare here two models, both with $\alpha_{\rm CE}=0.3$ --
which is consistent  with the results of  \cite{Zorotovic_2010} -- and
$n(q)=1$, but  with $\alpha_{\rm int}=0.0$ or  $\alpha_{\rm int}=0.2$,
respectively.  We have chosen these two particular models to highlight
the effects  of including  a fraction  of the  internal energy  of the
envelope that helps in the  ejection process.  The left-hand panels of
Fig.~\ref{f:lambdas}  display  the  results  for the  model  in  which
$\alpha_{\rm int}=0.0$,  while the right-hand  ones are for  the model
with $\alpha_{\rm int}=0.2$.   Systems that have experienced  a case B
CE episode  are displayed using green  dots, while blue dots  show the
WD+MS systems which  underwent a case C CE episode  and red dots those
in which a TPGAB CE episode took  place. As can be seen in this figure
for those models in which no internal energy is available to eject the
envelope the value of $\lambda$  remains practically constant and with
a relatively  small dispersion  which increases first  with increasing
Roche-lobe   radius,   until  it   reaches   a   maximum  at   $R_{\rm
L}\sim200\,R_{\sun}$, and  then decreases  again for larger  values of
$R_{\rm L}$ (see  the top-left panel of  Fig.~\ref{f:lambdas}). On the
other hand, when a moderate amount  of internal energy is available to
eject the  envelope we  find an overall  enhancement of  the resulting
values of  $\lambda$ (top-right panel of  Fig.~\ref{f:lambdas}).  This
was expected  since the  contribution of  the internal  energy becomes
more important  for more  extended envelopes, where  the gravitational
energy   becomes   smaller   and   the  envelope   is   less   tightly
bound. Moreover, this  enhancement is more noticeable  for the largest
values of  the Roche-lobe radius  at which  the CE episode  occurs. We
also find that the dispersion in the values of $\lambda$ increases for
wider systems. In this sense, we emphasize that the top left and right
panels Fig.~\ref{f:lambdas} actually show for which binary systems the
contributions of the internal energy are prominent. The progenitors of
systems  with He-core  WDs fill  their Roche-lobe  on the  first giant
branch where only a very small  amount of internal energy is stored in
the  envelope.   Thus, for  those  systems,  increasing the  value  of
$\alpha_{\rm int}$  does not lead  to an increased value  of $\lambda$
and  has virtually  no effect  on the  outcome of  CE evolution.   The
distributions of  primary ZAMS masses and  WD masses as a  function of
the Roche-lobe  radius is rather  similar for both models  (second and
third  panel from  top,  respectively). Finally,  the distribution  of
orbital  periods is  also very  similar in  both cases,  except for  a
population of long  period ($\ga 10$~days) PCEBs,  descending from the
initially  more  separated systems.   This  is  only observed  when  a
fraction  of  the  internal  energy  of the  envelope  is  taken  into
account. In summary, the only relevant differences between both models
are the distribution  of the values of $\lambda$ and  the existence of
systems  with  very  long  final   periods,  being  the  rest  of  the
distributions very similar.

\begin{figure*}[t]
\vspace{15.0cm}
\includegraphics{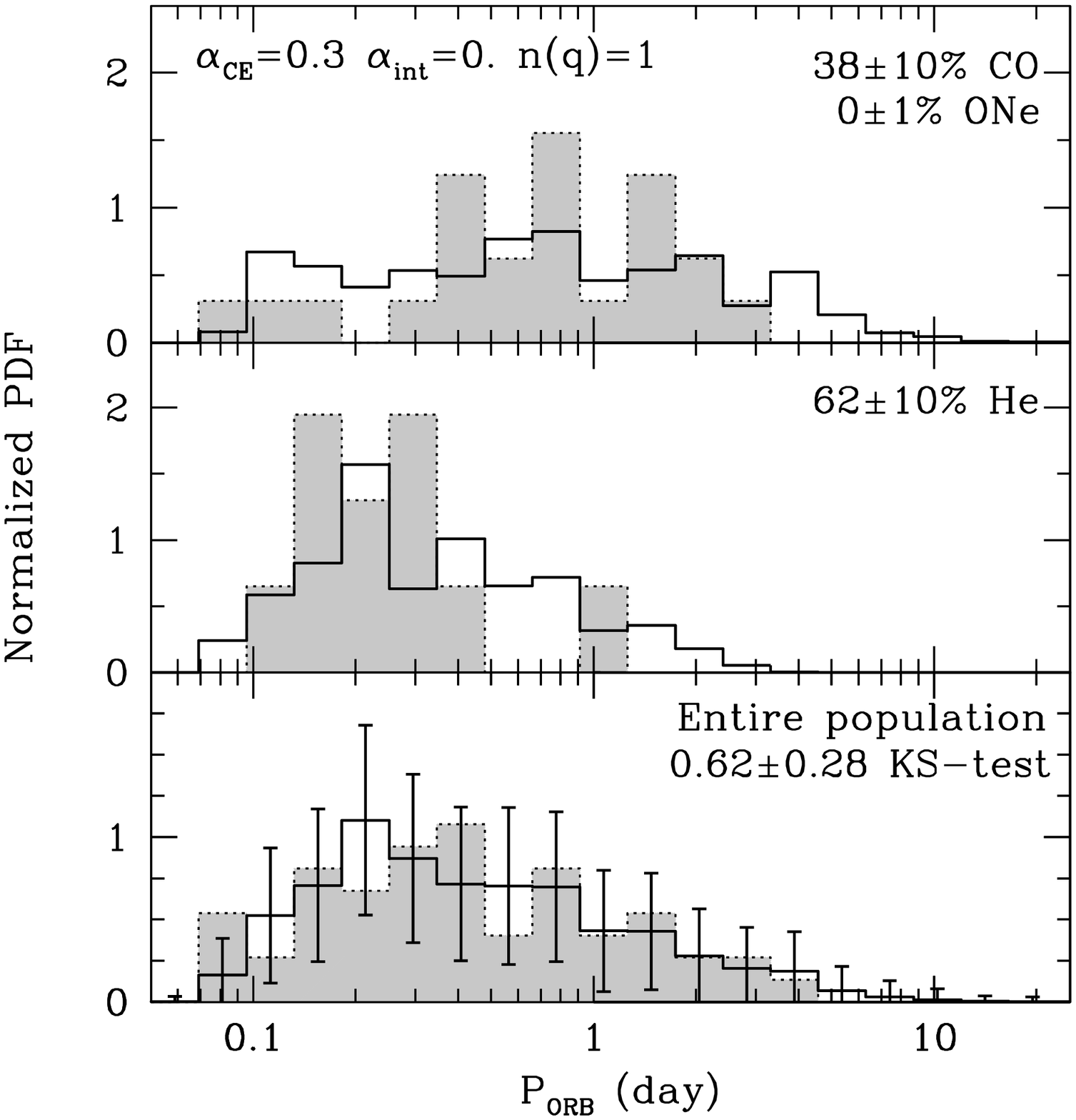} 
\includegraphics{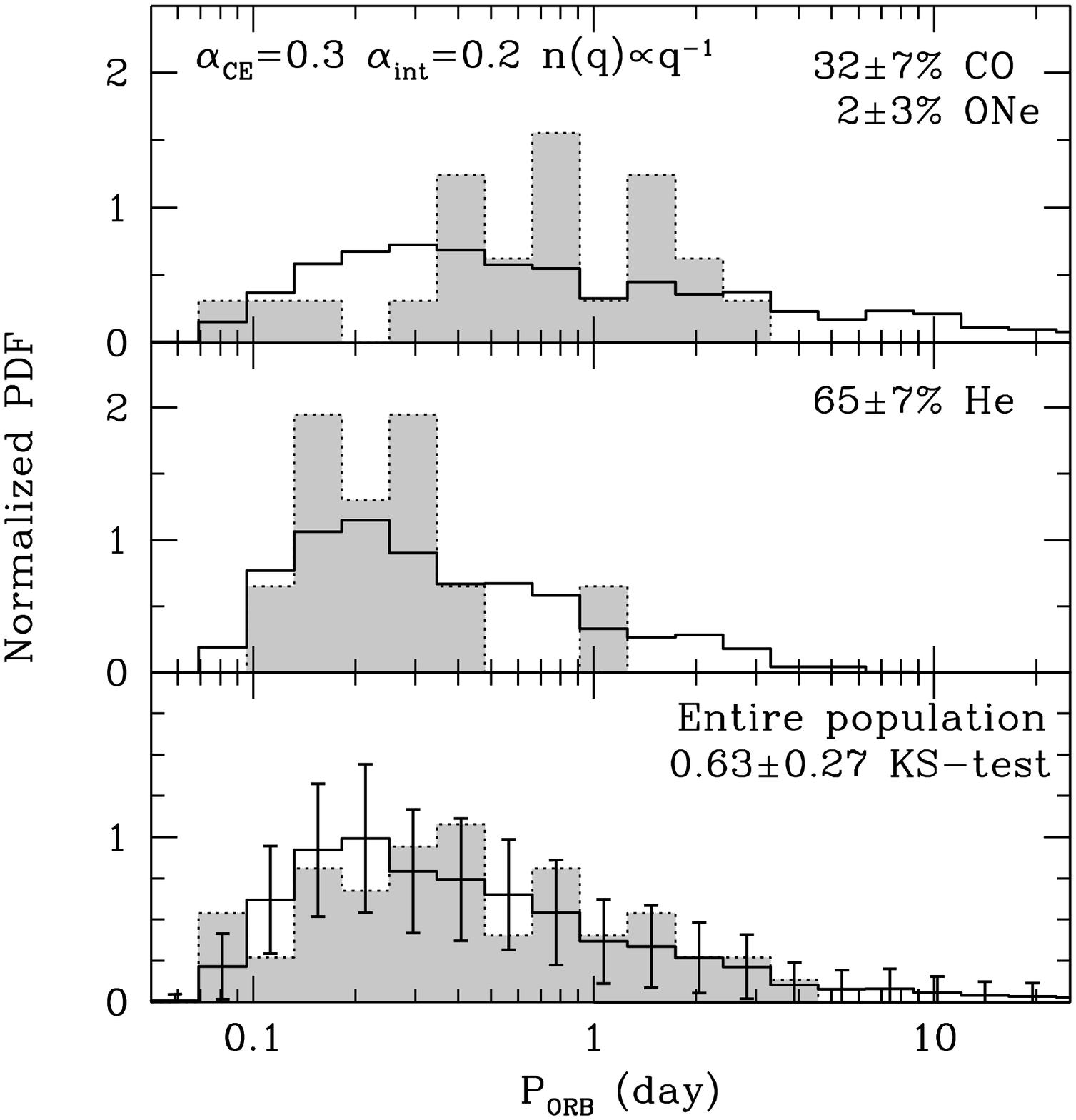}
\includegraphics{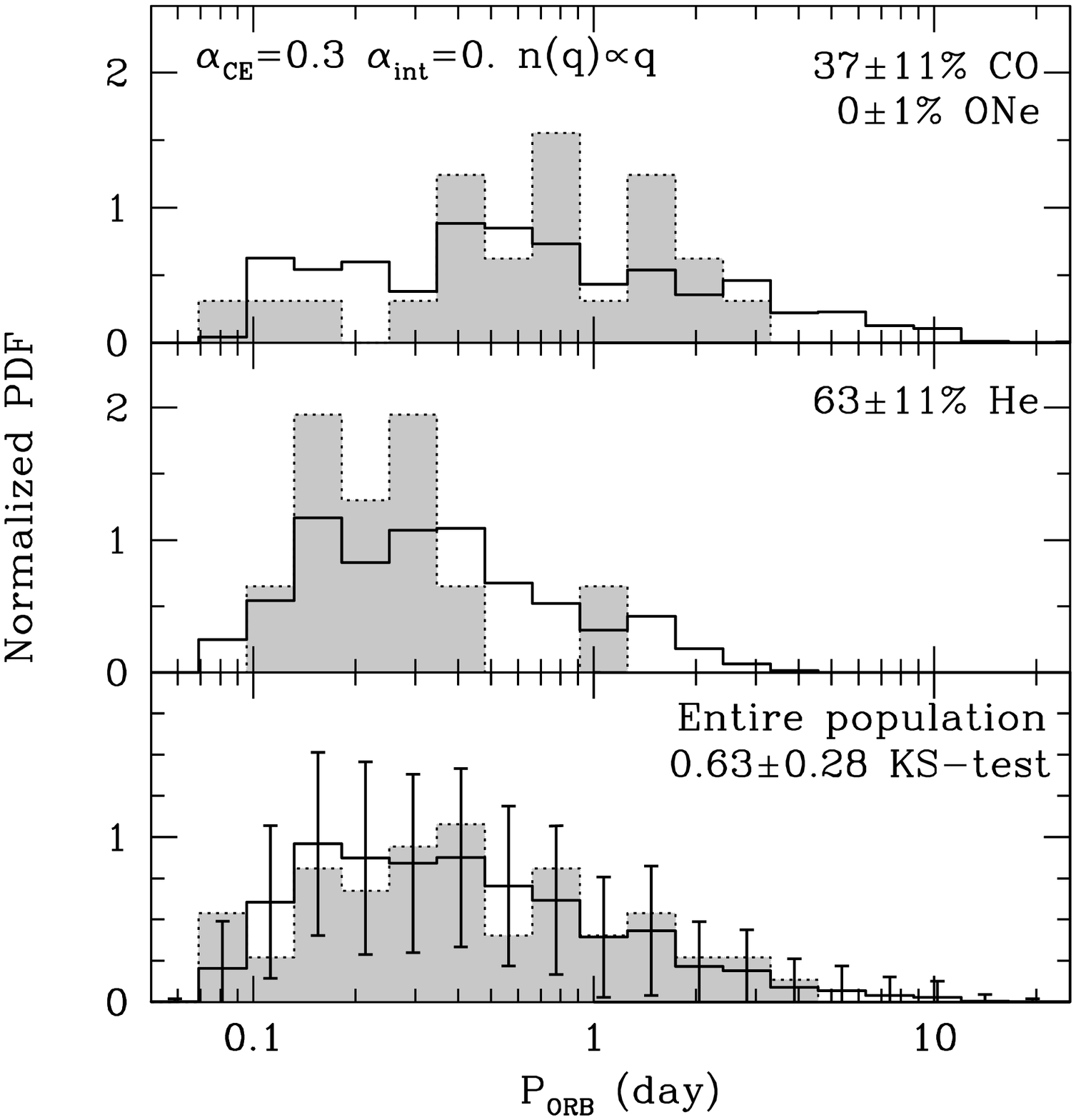} 
\includegraphics{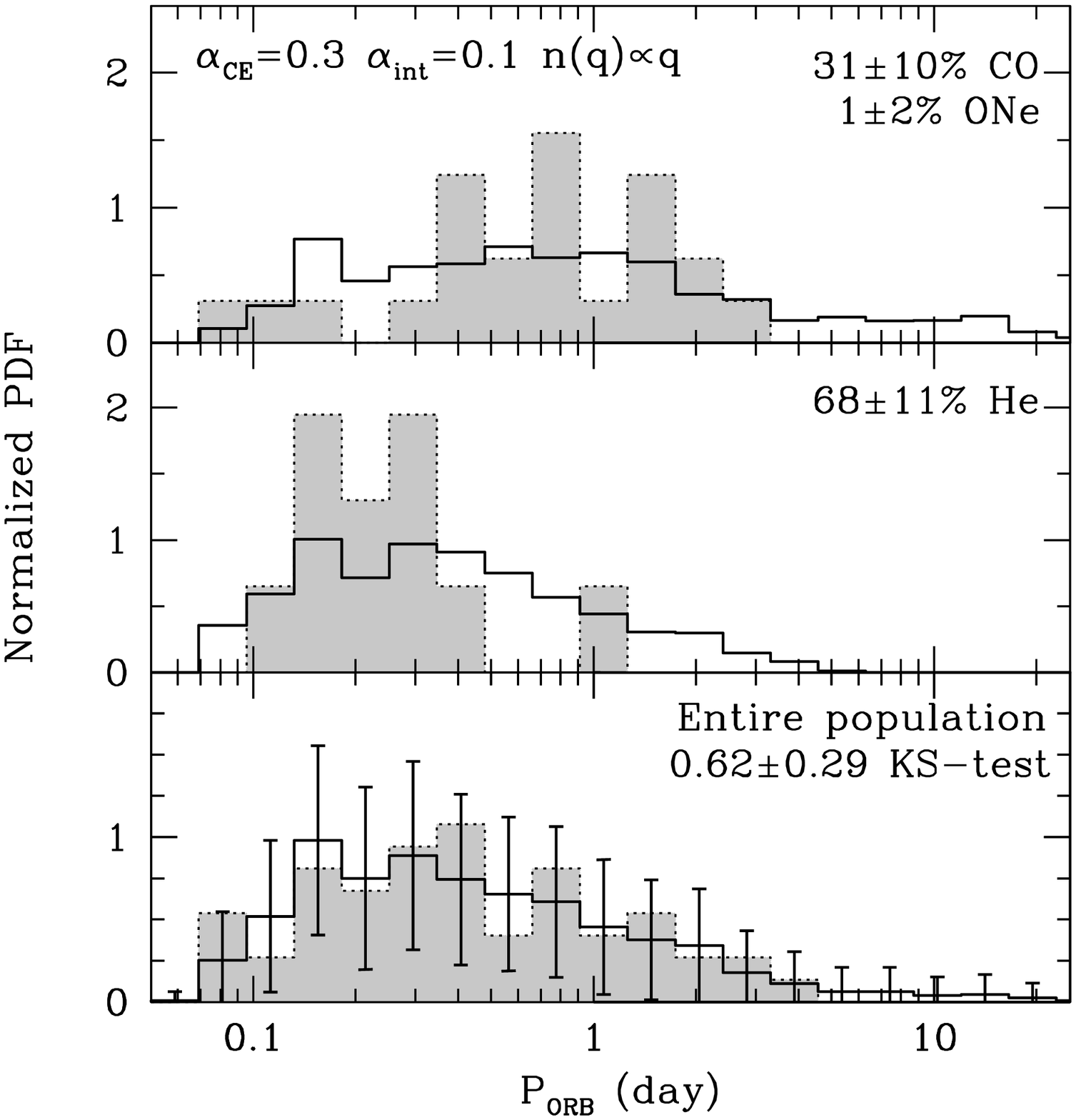} 
\caption{Period   histograms  (normalized   to  unit   area)  of   the
  distribution of  present-day WD+MS  PCEBs for  our four  best models
  (black line)  compared with  the observational  distribution (dotted
  line, gray histogram).}
\label{f:4phisto}
\end{figure*}

\begin{table*}[t]
\begin{center}
\caption{Percentage of systems  with He WDs and KS test  of the period
  distribution  for  our  models  with  $\alpha_{\rm  CE}\leq0.3$  and
  $\lambda$  properly  computed  for   each  system,  where  different
  fractions of internal energy are taken into account.
\label{t:kstest:variable}}
\tiny
\begin{tabular}{lccccccccc}
\hline
\hline
$n(q)$& &$\propto q^{-1}$ & & &$1$ & & &$\propto q$ & \\
\hline
&\multicolumn{9}{c}{$\alpha_{\rm int}=0.0$}\\
\cline{2-10}
$\alpha_{\rm CE}$ & 0.1 & 0.2 & 0.3 & 0.1 & 0.2 & 0.3 & 0.1 & 0.2 & 0.3 \\
He (\%) & $37\pm 13$ & $50\pm 14$ & $58\pm 8$ & $38\pm 15$ & $53\pm 14$ & $62\pm 10$ & $41\pm 15$ & $57\pm 12$ & $63\pm 11$ \\
KS & $0.20\pm 0.25$ & $0.38\pm 0.31$ & $0.49\pm 0.31$ & $0.35\pm 0.30$ & $0.52\pm 0.31$ & $0.62\pm 0.28$ & $0.37\pm 0.31$ & $0.54\pm 0.30$ & $0.63\pm 0.28$ \\
\hline
&\multicolumn{9}{c}{$\alpha_{\rm int}=0.1$}\\
\cline{2-10}
$\alpha_{\rm CE}$ & 0.1 & 0.2 & 0.3 & 0.1 & 0.2 & 0.3 & 0.1 & 0.2 & 0.3 \\
He (\%) & $48\pm 13$ & $55\pm 13$ & $64\pm 8$ & $47\pm 15$ & $56\pm 14$ & $65\pm 10$ & $52\pm 15$ & $63\pm 11$ & $68\pm 11$ \\
KS & $0.48\pm 0.31$ & $0.53\pm 0.30$ & $0.56\pm 0.29$ & $0.48\pm 0.31$ & $0.55\pm 0.30$ & $0.57\pm 0.30$ & $0.44\pm 0.31$ & $0.59\pm 0.29$ & $0.62\pm 0.29$ \\
\hline
&\multicolumn{9}{c}{$\alpha_{\rm int}=0.2$}\\
\cline{2-10}
$\alpha_{\rm CE}$ & 0.1 & 0.2 & 0.3 & 0.1 & 0.2 & 0.3 & 0.1 & 0.2 & 0.3 \\
He (\%) & --- & $59\pm 13$ & $65\pm 7$ & --- & $57\pm 17$ & $72\pm 10$ & --- & $64\pm 11$ & $70\pm 10$ \\
KS & --- & $0.55\pm 0.30$ & $0.63\pm 0.27$ & --- & $0.58\pm 0.29$ & $0.61\pm 0.29$ & --- & $0.58\pm 0.29$ & $0.62\pm 0.29$ \\
\hline
&\multicolumn{9}{c}{$\alpha_{\rm int}=0.3$}\\
\cline{2-10}
$\alpha_{\rm CE}$ & 0.1 & 0.2 & 0.3 & 0.1 & 0.2 & 0.3 & 0.1 & 0.2 & 0.3 \\
He (\%) & --- & --- & $69\pm 9$ & --- & --- & $70\pm 11$ & --- & --- & $71\pm 10$ \\
KS & --- & --- & $0.55\pm 0.30$ & --- & --- & $0.50\pm 0.31$ & --- & --- & $0.58\pm 0.30$ \\
\hline
\end{tabular}
\end{center}
\end{table*}

\subsection{The fraction of PCEBs containing He-WDs}
\label{s-periods}

One important and relatively robust value that can be derived from the
observed sample is  the fraction of PCEBs containing  He-core WDs.  We
therefore here  compare the  percentage of  WDs with  He cores  in the
final  sample of  our simulations  with that  of the  observed sample,
which is around $40\%$.

In table~\ref{t:kstest:fixed} we display the  percentage of He WDs, as
well as the results of a Kolmogorov-Smirnov (KS) test resulting from a
comparison   between  the   observed   and   the  theoretical   period
distributions  (we   will  describe  and   discuss  the  KS   test  in
Sect.~\ref{sec:KS}), for some of the  Monte Carlo simulations in which
a fixed  value of $\lambda=0.5$ was  adopted, for the three  IMRDs. We
emphasize  that for  the  sake of  conciseness in  this  table only  a
selected handful  of models  is shown. However,  the actual  number of
models analyzed is  much larger.  As can be seen,  a common feature of
the  synthetic distributions  is the  resulting large  fraction of  He
WDs. Specifically, the results displayed in table~\ref{t:kstest:fixed}
-- and those obtained from similar models not explicitly shown here --
show that only the models for which a small value of $\alpha_{\rm CE}$
is adopted produce the required  percentage of He WDs.  In particular,
in all the models in which  $\alpha_{\rm CE}$ is larger than $0.3$ the
fraction  of  WDs with  He  cores  is  significantly larger  than  the
observed value,  $36\pm 8\%$.  This  is true  for all three  IMRDs. We
think  that the  large fraction  of He  WDs found  in our  Monte Carlo
simulations  is  not a  weakness  of  the  models, but  a  potentially
interesting  feature that  deserves further  study. However,  we judge
that  this result  should  be taken  with some  caution,  as the  core
composition of the  synthetic WDs is set by  its evolutionary history,
and depends  on the adopted mass  limit between He and  C/O WDs. Also,
the observed fraction of He WDs  depends crucially on the error in the
mass  determinations of  the  of  WDs in  the  sample  of PCEBs  WD+MS
binaries   in  the   SDSS.   This   issue  was   explored  before   by
\cite{rebassa-mansergasetal11-1} --  see, for instance, their  Table 4
-- who  found  that  as  the  uncertainty in  the  mass  estimates  is
typically  $\sim 0.05-0.1  \, M_{\sun}$,  the theoretically  predicted
clear separation between He and C/O WDs at $M_{\rm WD}=0.5\, M_{\sun}$
is smeared out.  Hence, the real  observed fraction of He WDs in PCEBs
WD+MS systems  is still subject to  some uncertainty, and needs  to be
better determined, since it might be  possible that some of the He WDs
have instead C/O cores.

Table~\ref{t:kstest:variable} shows the same  results but for the case
in which  $\lambda$ is computed  for different values  of $\alpha_{\rm
int}$.  Again,  we do  this for several  values of  $\alpha_{\rm CE}$,
$\alpha_{\rm int}$ (with $\alpha_{\rm  int} \le \alpha_{\rm CE}$), and
for the three IMRDs. Based on  our previous results, we only show here
the results for our models  with $\alpha_{\rm CE}\le0.3$.  Once again,
the fraction of  WDs with He cores depends sensitively  on the adopted
value of $\alpha_{\rm CE}$, and also  a bit on $\alpha_{\rm int}$.  In
particular, as $\alpha_{\rm CE}$ is increased the percentage of He WDs
also increases, independently of the adopted IMRD.

\subsection{The orbital period distribution}
\label{sec:KS}

The parameter  of PCEBs that  can be  most accurately measured  is the
orbital  period. Thus,  comparing the  predicted and  observed orbital
period distribution is crucial. We  performed KS tests to estimate the
similitude    of   the    theoretical    and   observational    period
distributions. We restrict ourselves to models with $\alpha\leq0.3$ as
otherwise  the fraction  of PCEBs  containing He-core  WDs drastically
disagrees with  the observations  (see previous section).   All models
with $\alpha_{\rm CE}\leq\,0.3$ reproduce reasonably well the observed
orbital period distribution which  is indicated by KS-values exceeding
0.2.  This  means that  there are no  significant indications  for the
simulated and the observed distribution to be different. We obtain the
largest  KS-values  (exceeding  0.6)   for  models  with  $\alpha_{\rm
CE}=0.3$. In what  follows we describe the results  obtained for those
models that best fit the period distribution in some more detail.

\begin{table*}[t]
\begin{center}
\caption{Statistics for the best models.}
\label{t:best}
\begin{tabular}{lcccc}
\hline
\hline
\cline{1-5}
\cline{1-5}
\centering
$n(q)$ & $\propto q^{-1}$ & 1 & $\propto q$ &$\propto q$ \\
\cline{1-5}
$\alpha_{\rm CE}$ & 0.3 & 0.3 & 0.3 & 0.3 \\
$\alpha_{\rm int}$ & 0.2 & 0.0 & 0.0 & 0.1 \\
\cline{1-5} 
$N_{\rm WD+MS}$ & $42\pm 6$ & $24\pm 5$ & $19\pm 5$ & $20\pm 5$ \\
He ($\%$) & $65\pm 7$ & $61\pm 10$ & $63\pm 11$ & $68\pm 11$ \\
C/O ($\%$) & $32\pm 7$ & $38\pm 10$ & $37\pm 11$ & $31\pm 11$ \\
O/Ne ($\%$) & $ 3\pm 3$ & $ 0\pm 1$ & $ 0\pm 1$ & $ 1\pm 2$ \\
\hline
$\langle P\rangle$ (days) & $1.54\pm 7.20$ & $0.80\pm 1.32$ & $0.73\pm 1.33$ & $1.36\pm 7.16$ \\
$\langle P\rangle_{\rm He}$ (days) & $0.57\pm 0.74$ & $0.50\pm 0.50$ & $0.51\pm 0.52$ & $0.61\pm 0.67$ \\
$\langle P\rangle_{\rm C/O}$ (days) & $3.52\pm 12.24$ & $1.40\pm 1.24$ & $1.17\pm 1.75$ & $3.13\pm 12.27$ \\
$P_{\rm min}$ (days) & 0.049 & 0.067 & 0.068 & 0.067 \\
$P_{\rm max}$ (days) & 325 & 32 & 41 & 313 \\
\hline
\end{tabular}
\end{center}
\end{table*}

For the sake of conciseness we  only considered those models with a KS
value larger than  0.6, with a percentage of WD+MS  PCEBs with He-core
WDs smaller  than 70\%  -- see \cite{rebassa-mansergasetal11-1}  for a
detailed discussion of the percentage of  He WDs in WD+MS PCEBs in the
SDSS --  and a small  fraction ($<6\%$) of  O/Ne WDs --  in accordance
with the observed sample. Additionally,  we required that the selected
theoretical models had statistical properties  similar to those of the
observed sample of  WD+MS binaries.  These included  a similar average
period,  as well  as  maximum  and minimum  periods  of the  synthetic
binaries  after  applying the  successive  filters  in agreement  with
observations,  and  an assessment  of  the  morphology of  the  global
distribution of periods. Once these  criteria are employed we are left
with only four models.  The first model has $\alpha_{\rm int}=0.2$ and
$n(q)\propto  q^{-1}$, the  second  one has  $\alpha_{\rm int}=0$  and
$n(q)=1$, the third one has  $\alpha_{\rm int}=0$ and $n(q)\propto q$,
and finally the fourth one has $\alpha_{\rm int}=0.1$ and $n(q)\propto
q$.  Note that all the models correspond to a CE prescription in which
$\lambda$ is computed  for each binary.  Among these  four best models
there  is a  degeneracy between  the adopted  prescription for  the CE
phase and  the IMRD.  This  implies that on  the basis of  the present
observational data we cannot determine which is the IMRD.

In Figure~\ref{f:4phisto}  we compare  the distribution of  periods of
the present-day WD+MS PCEBs  simulated sample (white histograms, solid
lines) with the observational one (gray histograms, dotted lines).  We
show the  period distributions  for the entire  sample of  WD+MS PCEBs
(bottom  panel  of  each  figure)  but  also  separately  for  systems
containing  He  WDs   (middle  panels)  and  C/O  or   O/Ne  WDs  (top
panels).  From  the 40  systems  with  WD  mass determination  and  WD
temperature larger  than 12\,000 K described  in Sect.~\ref{s-sample},
we found  that six of them,  with WD masses close  to $0.5\,M_{\sun}$,
can contain either a  He WD or a C/O WD given their  WD mass error. Of
the 34  remaining systems, 11  contain a  He WD and  23 a C/O  or O/Ne
WD. These are the systems that  were considered for the middle and top
panels, respectively, while  the bottom panels contain  the 53 systems
with available periods.  In general, our Monte Carlo simulations agree
well  with  the  observational  period  distribution  for  the  entire
population. However, the still large  observational error bars and the
almost negligible differences between the different theoretical models
preclude from  drawing definite conclusions  of which of these  is the
best one. This is indicative  that the selection criteria dominate the
final observational distribution.  Nevertheless, a detailed inspection
of  Fig.~\ref{f:4phisto}  reveals  that  those  models  with  non-zero
internal energy present slightly extended tails in the long-period end
of the  distribution.  Even though  these tails possibly could  not be
statistically significant,  their mere existence provides  a hint that
these models do not describe  appropriately the ensemble properties of
the period distribution of WD+MS  PCEBs. Consequently, this compels us
to consider  as more convenient  those models  with a small  amount of
internal energy.

Table~\ref{t:best} contains the statistics  obtained for our four best
models. This table  also shows that those models  with non-zero values
of the internal  energy parameter have maximum periods  much larger (a
factor of $\sim 10$) than the  ones in which $\alpha_{\rm int}=0.0$ is
adopted, while the minimum periods remain nearly the same. The average
value for the  periods is therefore larger when we  include a fraction
of internal energy,  which is specially true for  systems containing a
C/O  or an  O/Ne WD.   Those  models in  which no  internal energy  is
available to  eject the CE fit  better the measured average  period of
the    observed     distribution    of    WD+MS     PCEBs    ($\langle
P\rangle=0.69$~days). It is  as well interesting to  remember that the
internal  energy   becomes  specially   important  for   more  evolved
primaries, which have  a more massive core (the future  WD) and a more
extended  envelope.   For  this  reason  those  simulations  in  which
$\alpha_{\rm int}\ne 0$  have an enhanced production  of WD+MS systems
with  an O/Ne  WD,  because it  becomes easier  for  these systems  to
survive a CE  phase due to this additional source  of energy.  This is
an important fact,  because in the observed sample there  are only two
WD+MS PCEBs  in which the resulting  WD has a mass  larger than $1.1\,
M_{\sun}$.  All in  all, we conclude that to account  for the ensemble
properties of the distribution of periods and the detection of a small
fraction of  WD+MS PCEBs with  very massive  WDs, the fraction  of the
internal energy available to eject the envelope must be small.

Finally, it is  worth mentioning that the average periods  for the two
sub-populations of  WDs with He  and C/O  or O/Ne cores,  are markedly
different, being that of WD+MS  systems with He core WDs significantly
smaller  than that  of  systems  with more  massive  WDs.  This is  in
agreement      with      the       observational      analysis      of
\citet{Zorotovic_2011}. If one separates He-core  and C/O or O/Ne core
systems, however, the number of  observed systems becomes too small to
separately compare model predictions and observations.

Finally, we  note that  although our population  synthesis simulations
reproduce with reasonable accuracy the global observed distribution of
orbital  periods,   this  is   not  the   case  when   the  individual
distributions for  He WDs and C/O  or O/Ne WDs are  considered, a fact
that  is somewhat  hidden by  the normalization  criteria employed  in
Fig.~\ref{f:4phisto}.  This  may be  indicative  of  missing piece  of
physics in the theoretical calculations or, as already mentioned, to a
not entirely reliable determination of  WD masses. However, the reader
should  keep in  mind  that the  theoretical  histograms presented  in
Fig.~\ref{f:4phisto} are  the result  of averaging  a large  number of
individual  Monte  Carlo  realizations.   In  a  typical  Monte  Carlo
realization  in which  $\sim  15$  WD+MS PCEBs  are  culled the  final
distributions are more  irregular, and would be more  similar to those
observationally  found.  Clearly,  additional  studies  are needed  to
clarify this issue.  Nevertheless, these studies are  beyond the scope
of the present paper.

\subsection{Period-mass distribution}
\label{sec:p-md}

\begin{figure*}[t]
\vspace{11.5cm}
\includegraphics{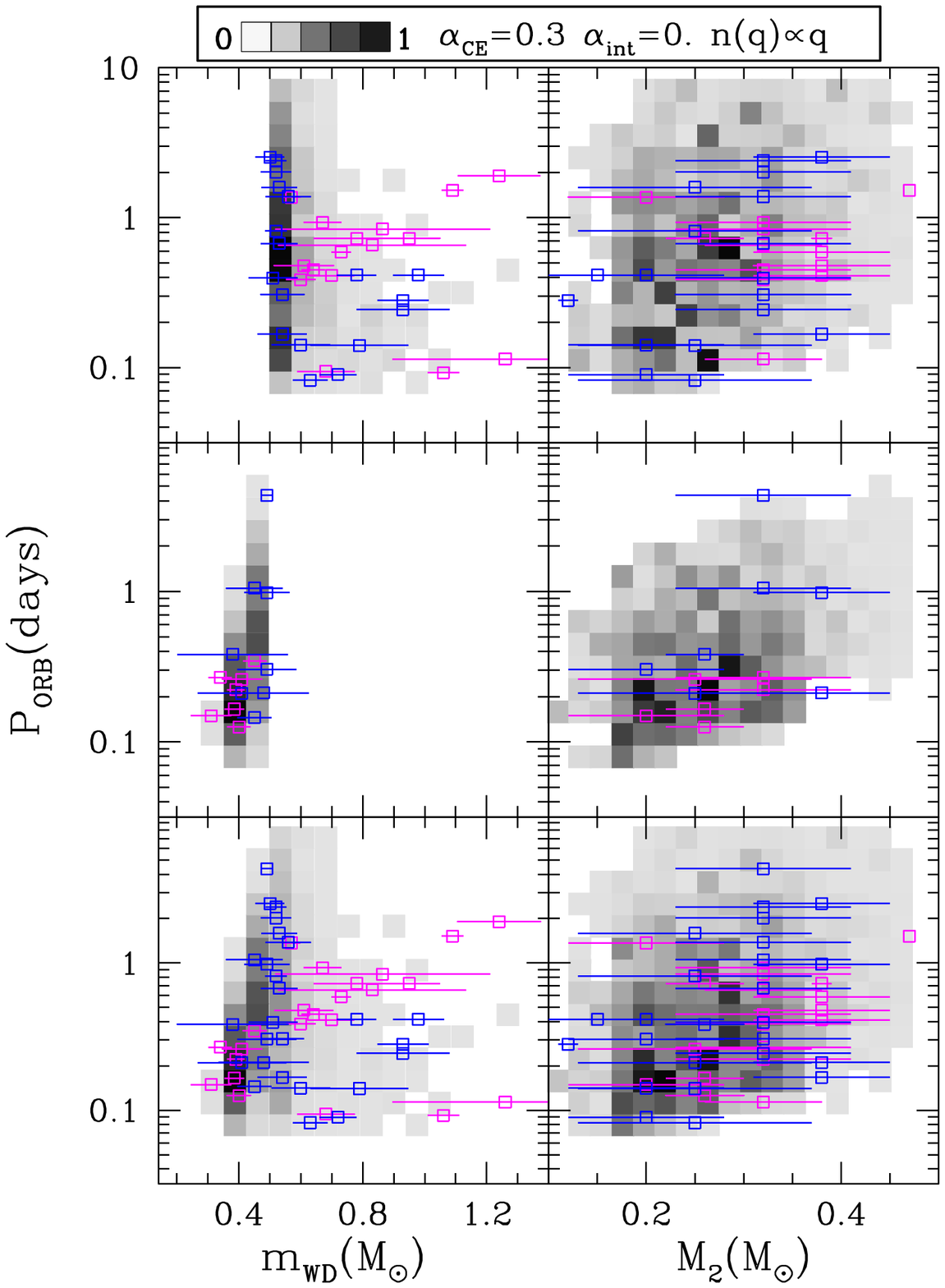} 
\includegraphics{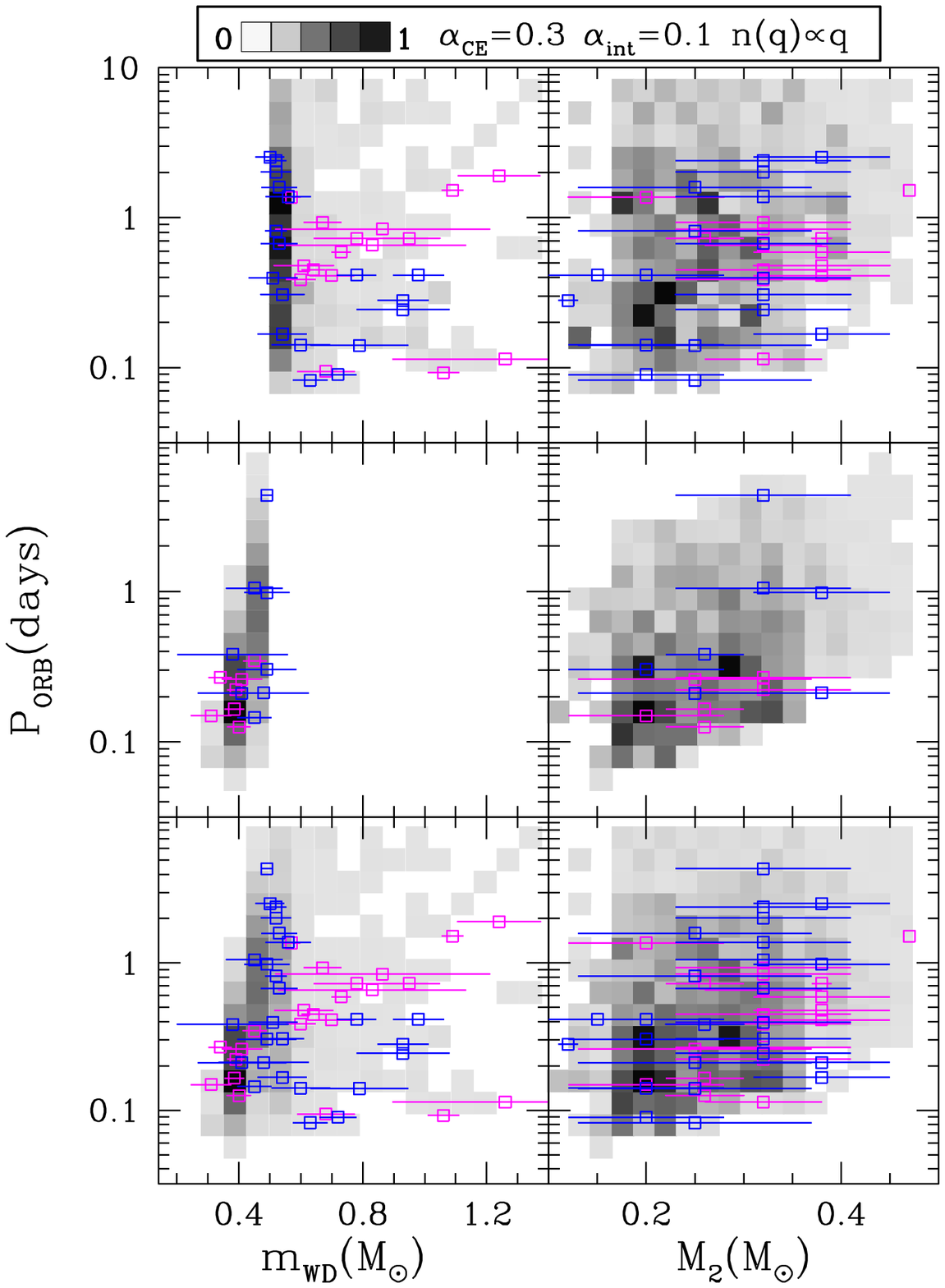}
\caption{Period-mass density  distribution of present-day  WD+MS PCEBs
  for  two of  our four  best models  (gray scale)  compared with  the
  observational  distribution (magenta  and blue  squares).  The  blue
  squares denote those systems for  which the effective temperature of
  the  WD  is   smaller  than  12,000  K,  in  which   case  the  mass
  determination of  the WD could  be problematic. The top  panels show
  the population  of WD+MS  PCEBs containing C/O  or O/Ne  WDs, middle
  panels are for systems containing a  He WD, and the bottom ones show
  the entire population of WD+MS PCEBs simulated.}
\label{f:density}
\end{figure*}

Figure~\ref{f:density}  shows  the  period-mass distributions  of  the
simulated PCEBs for two of  our best models ($\alpha_{\rm CE}=0.3$ and
$n(q)\propto q$)  without and  with the  inclusion of  internal energy
($\alpha_{\rm int}=  0.0$ and  $\alpha_{\rm int}=0.1$, left  and right
respectively). For each model the left panels show the distribution of
orbital periods as  a function of the WD mass,  while the right panels
show the same distribution as a function of the mass of the secondary.
As in Fig.~\ref{f:4phisto}  the top panels show  the sub-population of
systems containing  C/O or O/Ne WDs,  the middle panels those  with He
WDs, whilst  the bottom panels  show the distributions for  the entire
population. The magenta symbols correspond  to those WD+MS systems for
which the  uncertainty in the  mass determination  of the WD  is small
enough  to differentiate  between C/O  or O/Ne  WDs, and  He-core WDs,
while  the  blue  symbols  correspond  to  those  systems  which  have
effective temperatures smaller  than 12,000~K, in which  case the mass
determination  could be  problematic.  We  note that  in the  observed
sample there are 4 WDs with undetermined masses, 15 WDs with He cores,
none of which has an  effective temperature smaller than 12,000~K, and
34 systems hosting a C/O or O/Ne WD, of which 8 have effective smaller
than 12,000~K. Additionally, there are  4 binary systems for which the
mass of  the secondary remains  unknown.  Consequently, the  number of
observed  data  points  is  different   for  each  of  the  panels  of
Fig.~\ref{f:density}.

Clearly,   our  simulations   match  remarkably   well  the   observed
distribution of  WD+MS PCEBs (magenta  squares). It is  interesting to
note that the WD+MS binary systems that contain a He WD (middle panels
of  Fig.~\ref{f:density}) occupy  a  narrow strip  in  WD masses  and,
moreover,   the    periods   of    these   systems    cluster   around
0.2--0.3~days. All this is in  excellent agreement with the properties
of the observed  sub-population of WD+MS PCEBs with He  WDs. For those
WD+MS   binaries  containing   C/O  or   O/Ne  WDs   (top  panels   of
Fig.~\ref{f:density}) the  distribution of  WD masses  is considerably
broader, and most of the WD masses of our synthetic sub-population are
below $1.1\,  M_{\sun}$, and  thus are C/O  WDs. Our  simulations also
predict that WD+MS PCEBs containing  an O/Ne WD are possible, although
these systems  should be  rare, specially when  no internal  energy is
included.   This is  again consistent  with the  observed sample,
where only  2 systems contain an  O/Ne WD. The periods  of WD+MS PCEBs
with C/O  or O/Ne WDs also  span a larger range,  with typical periods
ranging from $\le  0.1$ to $\sim 4$ days, also  in good agreement with
the observations. When  all the WD+MS PCEBs with  available period and
masses are considered (bottom panels)  the agreement with the observed
distribution is excellent.

\section{Conclusions}
\label{s-concl}

In  this  paper  we  presented  a comprehensive  set  of  Monte  Carlo
simulations  of  the population  of  WD+MS  PCEBs  in the  SDSS.   Our
simulations  encompass  a very  broad  range  of possible  situations,
including three  IMRDs, different  prescriptions for the  treatment of
the CE episode, and of the parameters controlling the tidally enhanced
mass loss  during this phase. In  our simulations we included  all the
known systematic  observational biases. We  found that the  color cuts
reduce considerably the initial sample,  and that typically only $\sim
7\%$ of  the simulated  WD+MS PCEBs  survive the  cuts. The  number of
surviving   systems  is   further  reduced   when  the   spectroscopic
completeness filter is applied, leaving only $\sim 3\%$ of the systems
that previously survived the color cuts. The intrinsic binary bias and
the period filter additionally reduce  the total size of the simulated
samples, resulting  in total sample  sizes which  are of the  order of
$\sim 0.1\%$  of the initial  one.  All  in all, our  simulations show
that, given  the actual observational  capabilities, we are  probing a
very limited number of WD+MS  PCEBs, and that the observed sample
suffers  from poor  statistics.  This  prevents from  drawing definite
conclusions about the overall properties of the WD+MS PCEB population,
despite the huge  observational efforts done so  far. Additionally, we
also find  that the  population of  WD+MS PCEBs  containing He  WDs is
over-represented within SDSS due to selection effects.
 
Nevertheless,  a comparison  of our  population synthesis  simulations
with the  complete sample of  PCEBs currently available allowed  us to
draw some interesting conclusions, although we emphasize that to reach
physically  sound  conclusions the  theoretical  results  can only  be
compared  with  observations  once  all  the  observational  selection
effects  are properly  taken into  account.   Thus, in  this paper  we
simulated for  the first  time the entire  process of  discovery, PCEB
identification, and  orbital period determination of  PCEBs discovered
by SDSS and compared model  predictions and observations.  Our results
can be summarized as follows:

\begin{itemize}

\item Even for small values of the mass loss enhancement parameter the
  percentage of  He WDs  is at odds  with that  observationally found.
  Small values of  this parameter agree better  with the observational
  data set.

\item A small value of the CE efficiency ($\alpha_{\mathrm{CE}} \leq\,
  0.3$)  is  required to  reproduce  the  observed fraction  of  PCEBs
  containing He-core WDs. 

\item An  interesting feature of  our synthetic distributions  is also
  the resulting large fraction of He WDs in several of the theoretical
  distributions. Even our best-fit models  have large He WD fractions,
  although  they  agree  within  the  error  bars  with  the  observed
  distribution.  We judge that this issue is a potentially interesting
  feature that might  be real. However, the existence  of this feature
  deserves further study, from  both the theoretical and observational
  sides.

\item  Models with  a variable  binding energy  parameter seem  to fit
  better the observed distribution of periods than models in which the
  binding energy parameter is assumed to be constant.

\item Our  results also show  that large values of  $\alpha_{\rm int}$
  are ruled out by the  observations, although the ensemble properties
  of the  population of WD+MS PCEBs  do not allow us  to discard small
  values of $\alpha_{\rm int}$, say smaller than 0.2, approximately.

\item We have  also compared the distribution of orbital  periods as a
  function  of  the  mass  and   find  excellent  agreement  with  the
  observational  data.  Our simulations  can  not  only reproduce  the
  distribution  of  orbital  periods,  but also  the  observed  period
  distribution as a function of the mass of the WD if small values for
  the  CE efficiencies  and  a detailed  prescription  of the  binding
  energy parameter are assumed.

\end{itemize} 

Finally,  we note  that the  present analysis  suffers from  the still
scarce  number  of  WD+MS  PCEBs  that  have  been  identified  in  an
homogenous way.  This  prevents us to draw  more definite conclusions.
However, evidence for small CE efficencies is growing.


\begin{acknowledgements}  
This research was supported by AGAUR, by MCINN grant AYA2011–23102, by
the European  Union FEDER  funds, by the  ESF EUROGENESIS  project, by
AECI grant A/023687/09, by Fondecyt (grants 3110049 and 3130559).  MRS
thanks the Millennium Science Initiative, Chilean Ministry of Economy,
Nucleus P10-022-F.  Part of the  the research leading to these results
has  received funding  from the  European Research  Council under  the
European  Union's Seventh  Framework  Programme  (FP/2007-2013) /  ERC
Grant Agreement  n. 320964 (WDTracer).   BTG was supported in  part by
the UK=92s  Science and Technology Facilities  Council (ST/I001719/1).
ARM  acknowledges  financial  support from  the  Postdoctoral  Science
Foundation of China (grant 2013M530470) and from the Research Fund for
International  Young  Scientists  by   the  National  Natural  Science
Foundation of China (grant 11350110496).
\end{acknowledgements}


\bibliographystyle{aa}
\bibliography{wdms}

\end{document}